\documentclass[usenatbib,usegraphicx]{mn2e}
\newcommand{\Ledd}{$L/L_{\rm Edd}$}
\newcommand{\Mbh}{$M_{\rm BH}$}
\newcommand{\Hb}{H$\rm\beta$}
\newcommand{\Ox}{[\mbox{O\,{\sc iii}}]}
\newcommand{\C}{\mbox{C\,{\sc iv}}}
\newcommand{\Fe}{\mbox{Fe\,{\sc ii}}}
\newcommand{\He}{\mbox{He\,{\sc ii}}}
\newcommand{\Mg}{\mbox{Mg\,{\sc ii}}}
\newcommand{\rS}{$r_{\rm S}$}
\newcommand{\aox}{$\alpha_{\rm ox}$}
\newcommand{\aouv}{$\alpha_{\rm o,UV}$}
\newcommand{\hst}{{\it HST}}
\newcommand{\iue}{{\it IUE}}
\newcommand{\fnrepeat}[1]{$^{\ref{#1}}$}

\title[What controls the \C\ profile in AGN?]{What controls the \C\ line profile in active galactic
nuclei?}
\author[A. Baskin and A. Laor]{Alexei Baskin\thanks{E-mail: alexei@physics.technion.ac.il (AB); laor@physics.technion.ac.il (AL)} and Ari Laor\footnotemark[1] \\
Physics Department, Technion, Haifa~32000, Israel}
\begin{document}
\maketitle
\begin{abstract}
The high ionization lines in active galactic nuclei (AGN), such as
\C, tend to be blueshifted with respect to the lower ionization
lines, such as \Hb, and often show a strong blue excess asymmetry
not seen in the low ionization lines. There is accumulating evidence
that the \Hb\ profile is dominated by gravity, and thus
provides a useful estimate of the black hole mass in AGN. The shift
and asymmetry commonly seen in \C\ suggest that non gravitational
effects, such as obscuration and radiation pressure, may affect
the line profile. We explore the relation between the \Hb\ and \C\
profiles using UV spectra available for 81 of the 87 $z\le 0.5$ PG
quasars in the Boroson \& Green (BG92) sample. We find the
following: (1) Narrow \C\ lines (FWHM~$< 2000$~km~s$^{-1}$) are
rare ($\sim 2$ per cent occurrence rate) compared to narrow \Hb\
lines ($\sim 20$ per cent). (2) In most objects where the \Hb\
FWHM~$<4000$~km~s$^{-1}$ the \C\ line is broader than \Hb, but
the reverse is true when the \Hb\ FWHM~$>4000$~km~s$^{-1}$.
This argues against the
view that \C\ generally originates closer to the center, compared to \Hb.
 (3) \C\ appears to provide a
significantly less accurate, and possibly biased estimate
of the black hole mass in
AGN, compared with \Hb. (4) All objects where \C\ is strongly
blueshifted and asymmetric have a high \Ledd, but the reverse is not true.
This suggests that a high \Ledd\ is a necessary but not sufficient
condition for generating a blueshifted asymmetric \C\ emission.
(5) We also find indications for dust
reddening and scattering in `normal' AGN. In particular, PG
quasars with a redder optical-UV continuum slope show weaker \C\
emission, stronger \C\ absorption, and a higher optical continuum
polarization.

\end{abstract}

\section{INTRODUCTION}
Systematic differences between the broad low and high ionization
emission line profiles in the broad line region (BLR) of active
galactic nuclei (AGN) were first noted by Gaskell (1982) and
Wilkes (1984; 1986), and later confirmed by various other studies
(e.g. Espey et al. 1989; Corbin 1990; Steidel \& Sargent 1991;
Tytler \& Fan 1992; Laor et al. 1994; Wills et al. 1995; Sulentic
et al. 1995; Baldwin et al. 1996;  Corbin \& Boroson 1996;
Marziani et al. 1996; McIntosh et~al. 1999; Vanden Berk et al.
2001; Richards et al. 2002). The peak of the higher ionization
lines (e.g. \C) tend to be blueshifted with respect to the peak of
the low ionization lines (e.g. \Hb), and they often show an
asymmetric profile with a large excess emission on their blue
side, not seen in low ionization lines. These differences suggest
that the low and high ionization lines originate in physically
distinct components in the BLR, as suggested earlier based on
photoionization model fits to the emission line ratios (e.g.
Baldwin \& Netzer 1978; Collin-Souffrin, Dumont \& Tully 1982;
Wills, Netzer \& Wills 1985; Snedden \& Gaskell 2004). The origin
of the profile differences is not established yet, but a plausible
interpretation is that the blueshift is induced by the combined
effect of an outflow and obscuration of the high ionization
component versus a virialized unobscured velocity distribution for
the low ionization component. The range of amplitudes of the
observed profile differences could either be a pure viewing aspect
effect in objects which are otherwise intrinsically similar (e.g.
Richards et al. 2002), or it could be related to intrinsic
differences in the dynamics of the high ionization component
(Baldwin et al. 1996; Laor et al. 1997a; Peterson et al. 2000;
Leighly 2004a).

There are now good observational indications that the \Hb\ width
is dominated by gravity. Black hole mass, \Mbh, estimates based on
the \Hb\ width give values consistent with the correlation of
\Mbh\ and host galaxy bulge luminosity and stellar velocity
dispersion $\sigma_*$ seen in nearby inactive galaxies (Laor 1998;
Gebhardt et al. 2000; Ferrarese et al. 2001). The \Hb\ region is
generally not available in objects with $z>1$, and recent studies
of \Mbh\ in $z>1$ AGN resorted to using the \C\ width instead of
\Hb\ (Vestergaard 2002, 2004; Netzer 2003; Corbett et al. 2003;
Warner, Hamann \& Dietrich 2003, 2004; Dietrich \& Hamann 2004).
Although efforts were made to calibrate the \C\ based  \Mbh\
estimates using the \Hb\ results, the differences in line profiles
raise the possibility that the \C\ profile may be affected by non
gravitational forces, which would introduce errors in the \Mbh\
estimates.

The purpose of this paper is to obtain observationally
based clues for the physical origin of the differences between the
\C\ and \Hb\ profiles in AGN. We make a systematic study of the
\C\ versus \Hb\ profiles in a nearly (93 per cent) complete sample
of low $z$ bright and well studied AGN. We have studied the
\C\ equivalent width (EW) distribution in this sample in an
earlier paper (Baskin \& Laor 2004, hereafter BL04). Here we
extend the BL04 study by measuring the various \C\ profile
parameters, and exploring their correlations with the \Hb\
parameters,  as taken from the extensive analysis of Boroson \&
Green (1992, hereafter BG92). In Section 2 we describe the
analysis of the optical and UV spectra, and how the emission line
parameters were measured. In Section 3 we describe the results of
the correlation analysis and their implications, and the main
conclusions are given in Section 4.

\section{THE DATA ANALYSIS}

\subsection{The sample}
For the purpose of this analysis we use the BG92 sample which
includes the 87 $z\le0.5$ AGN from the Bright Quasars Survey (BQS;
Schmidt \& Green 1983). This sample extends in luminosity from
Seyfert galaxies with $\nu L_{\nu}=3.3\times 10^{43}$~erg~s$^{-1}$
(calculated at rest frame 3000~\AA\ using the continuum fluxes in
Neugebauer et al. 1987,
assuming $H_0=80$~km~s$^{-1}$~Mpc$^{-1}$, $\Omega_0=1.0$), to
luminous quasars at $\nu L_{\nu}=1.4\times 10^{46}$~erg~s$^{-1}$.
This is a complete
and well defined sample, selected based on (blue) color and (point
like) morphology, independently of the emission line strengths and
profiles. It is also the most thoroughly explored sample of AGN,
with a wealth of high quality data at most wave bands, thus making
it an optimal sample for comparison of the \C\ and \Hb\ profiles
and their relation to other emission properties.

\subsection{The optical spectra}
The optical observations of the \Hb\ region of the 87 objects are
described in BG92, and were kindly made available to us  by T.
Boroson (private communication).
The rest-frame wavelengths were calculated using redshifts
determined from the peak of the \Ox\ $\lambda$5007 line
(see Table 1), kindly provided to
us by T. Boroson. The I Zw 1 \Fe\ template, also
provided by T. Boroson, was used to subtract the \Fe\ lines from
the spectra. The template was broadened by convolving with a
Gaussian, scaled by multiplying by a constant, and then subtracted
from the optical spectrum of each object. The best-fit residual
spectrum was searched by eye, while varying the broadening and the
absolute flux, based on the criterion of obtaining a featureless
continuum between H$\rm\gamma$ at 4340~\AA\ and \He\ $\lambda$4686
(BG92). The template was shifted by $\sim 100$ km~s$^{-1}$ for
19 objects, yielding a better match to the \Fe\ emission of
those objects.

\subsubsection{The absolute flux scale}
The BG92 spectra were obtained through a relatively narrow slit
which may degrade the absolute spectrophotometric accuracy. BG92
therefore used other spectra (mostly from Neugebauer et al. 1987)
to obtain the absolute flux level at rest frame 5500~\AA, which
they converted to an absolute {\it V} magnitude, $M_{\it V}$. We
used the cosmology assumed by BG92
($H_0=50$~km~s$^{-1}$~Mpc$^{-1}$, $\Omega_0=0.2$), to derive back
the observed flux density they used at rest frame 5500~\AA. We
then measured the corresponding observed flux density in the BG92
spectra. The ratio of the two fluxes defines a correction factor
(see Table 1), which we applied to the BG92 spectra in order to
get a better spectrophotometric accuracy. Since the BG92 spectra
are not corrected for Galactic extinction, while $M_{\it V}$
are corrected, the correction factor also corrects
for the Galactic extinction (the differential extinction across
BG92 optical spectra is negligible). This process proved
essential as otherwise the `raw' BG92 spectra yielded optical to
UV spectral slopes which were far too flat (mean $f_{\nu}$ slope
of $\sim 0$), while the corrected spectra yielded more plausible
values ($\sim -0.5$, Table 1), consistent with the known mean
quasar spectral shape.

\subsubsection{The \Hb\ profile}
In order to find the BLR \Hb\ profiles we subtracted the local
continuum and the contribution from the narrow line region (NLR).
A local power-law continuum was fitted to each spectrum between
$\sim 4600$~\AA\ and $\sim 5100$~\AA, the continuum flux density
at $\lambda$4861 was calculated from the fit (see Table 1), and
the power-law fit was subtracted from the spectrum. The \Ox\
$\lambda$5007 line profile was used to subtract the narrow
component of \Hb. No velocity shift was allowed between the NLR
components of \Hb\ and \Ox\ $\lambda$5007, and the only free
parameter was the narrow component flux. We subtracted the maximum
allowed narrow component which does not produce a dip in the \Hb\
profile, as determined by eye inspection. The narrow component of
\Hb\ generally contributes $<3$ per cent of the total line flux,
as noted by BG92, but it is significantly stronger in some of the
objects. The EW of the narrow \Hb\ component is listed
for each object in Table 1 (column 17). The \Hb\ profiles of all
objects are presented in Figure 1. However, for the purpose of the
correlation analysis we use the \Hb\ profile parameters as
reported by BG92\footnote{The \Hb~FWHM of PG~1307+085 was
corrected to 4190~km~s$^{-1}$, and of PG~2304+042 to
6500~km~s$^{-1}$.}, with the addition of the following two
parameters measured here, the continuum flux density
and the line flux density at 4861~\AA.

\subsection{The UV spectra}
A complete description of the analysis of the UV spectra is
provided in BL04, and is briefly reviewed here. Archival UV
spectra of the \C\ region  are available for 85 of the 87 BG92
objects (PG~1354+213 and PG~2233+134 had no archival UV
spectra). The \hst\ archives contain UV spectra of 47 of the
objects, which were obtained by the Faint Object Spectrograph
(FOS); the UV spectra for the remaining 38 objects with no \hst\
spectra were obtained from the \iue\ archives (see Table~1). An
average spectrum, weighted by the S/N ratio, was calculated when
more than one archival spectrum was available. Three of the
archival spectra did not have a sufficient S/N to measure the \C\
line (of PG~0934+013, PG~1004+130 and PG~1448+273),
and in one object \C\ is heavily absorbed (PG~1700+518, e.g.~Laor \& Brandt
2002), leaving a sample of 81 objects for the analysis.
We corrected the spectra for Galactic reddening using the
{\it E(B$-$V)} values from Schlegel, Finkbeiner \& Davis (1998, as
listed in the NASA/IPAC Extragalactic Database), and the reddening
law of Seaton (1979). Absolute wavelength calibration of the FOS
spectra was carried out using interstellar absorption lines, when
available (see Laor \& Brandt 2002).

\subsection{The \C\ profile}
A local power-law continuum was fit to each spectrum between $\sim
1470$~\AA\ and $\sim 1620$~\AA, and a narrow (\Ox\ like)
\C\ component was
subtracted following the procedure described above for \Hb. This
narrow component is typically very weak to non detectable,
as found in earlier studies (Wills et al. 1993b; Corbin \& Boroson
1996; Vestergaard 2002). The subtraction of a narrow  \C\ component
was performed to insure identical analysis of the \Hb\ and \C\
profiles. The resulting broad \C\ line emission was
fit as a sum of three Gaussians, using the procedure described in
Laor et al. (1994, section 3 and the appendix there). Wavelength
regions suspected to be affected by intrinsic or Galactic
absorption were excluded from the fit (Laor \& Brandt 2002). The
purpose of the three Gaussians fit is not to decompose the line to
possible components, as such a decomposition is not unique, but
rather to obtain a smooth realization of the line profile, which
is likely to yield more accurate values for the line profile
parameters.

Figure 1 presents all the measured \C\ profiles. For each object
we show both the smooth fit and the data points. To clarify the
presentation we rebinned the UV data, where the \hst\ spectra were
rebinned by a factor of four (each four data points were replaced
by their average), and the \iue\ spectra by a factor of two. The
\Hb\ profiles are plots of the data, and again to clarify the
presentation we smoothed the \Hb\ profile by convolving the data
with a $\sigma=1.78$ Gaussian (FWHM~$\sim 4$ pixels, which produces
a negligible line broadening).

\begin{figure*}
\includegraphics[width=110mm,angle=-90]{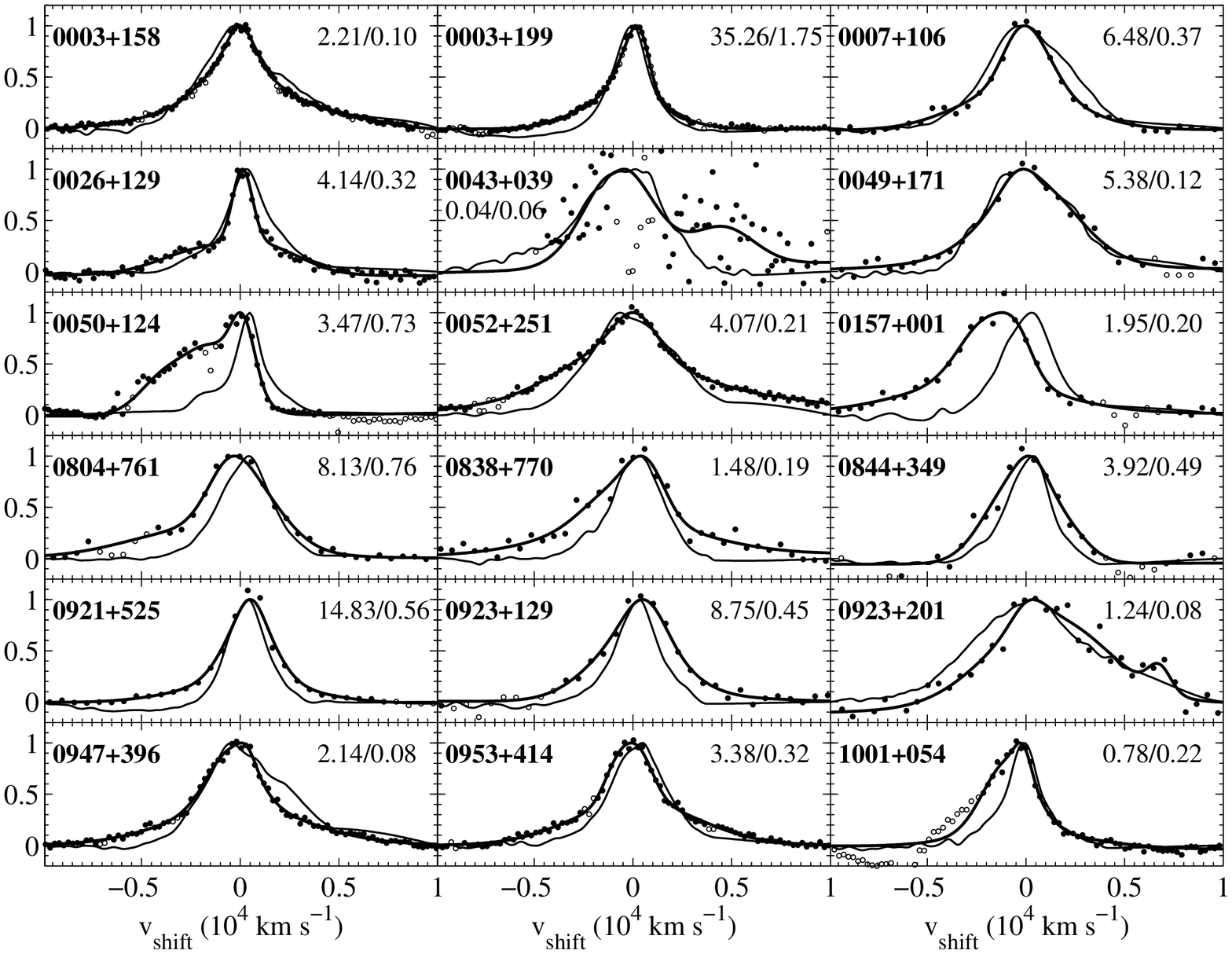}
\includegraphics[width=110mm,angle=-90]{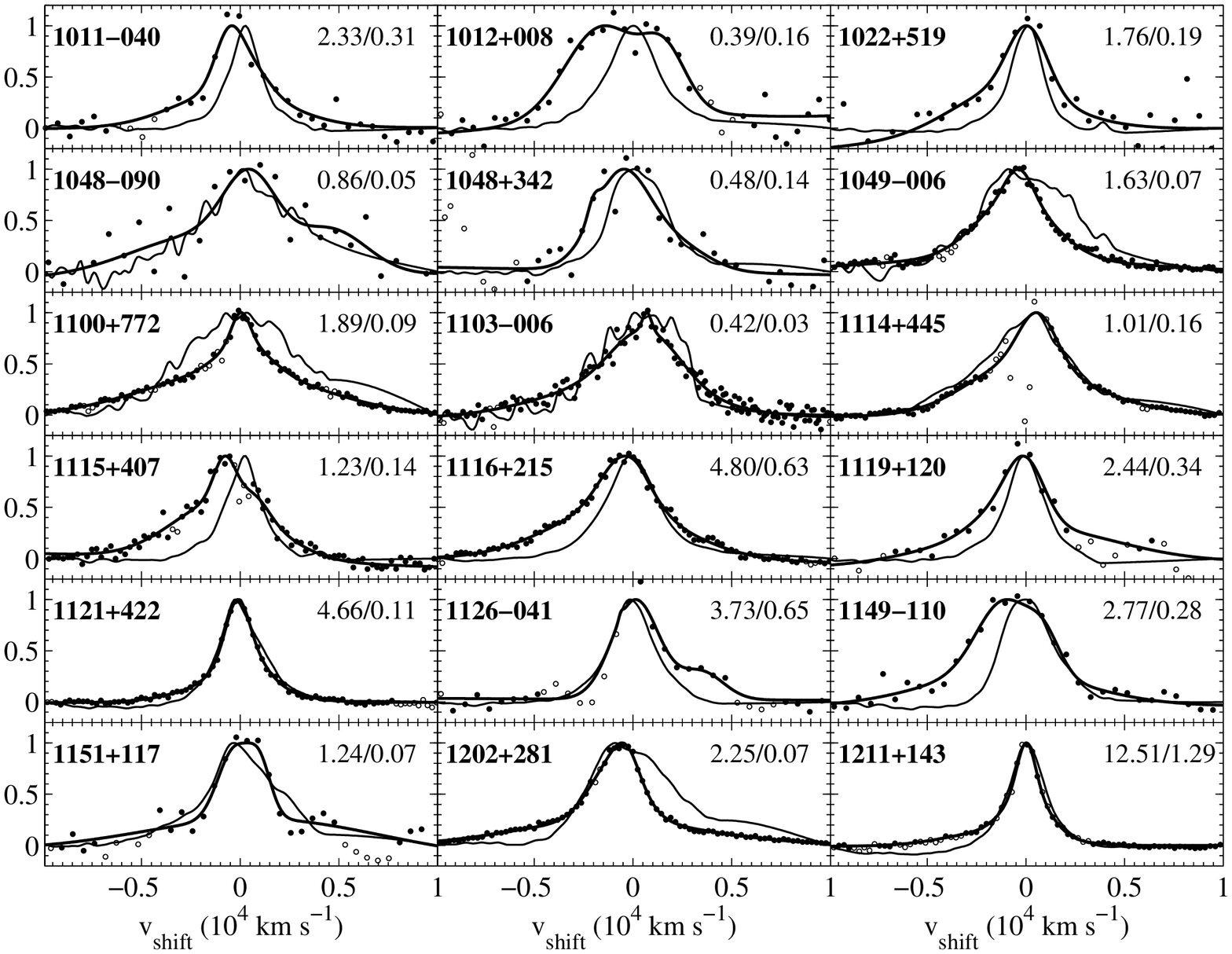}
\caption{A comparison of the \Hb\ and \C\ profiles. The name of
the object is indicated in each panel (in bold). The thin line is
the smoothed \Hb\ profile, and the thick line is the fit to the
\C\ profile (see text). Both profiles are normalized by their peak
flux density. The peak flux densities for \C/\Hb\ are listed in
each panel in units of
$10^{-14}$~erg~cm$^{-2}$~s$^{-1}$~\AA$^{-1}$. The circles indicate
the binned \C\ data points. Filled circles are data points used in
the fitting procedure, and empty circles are data points which
were excluded due to possible intrinsic or Galactic absorption
(Laor \& Brandt 2002). Note the remarkable profile similarity in
some objects (e.g. PG~0003+158; PG~1121+422; PG~1416$-$129), and
large differences in others (e.g. PG~0157+001; PG~1259+593;
PG~1543+489).}
\end{figure*}
\setcounter{figure}{0}
\begin{figure*}
\includegraphics[width=110mm,angle=-90]{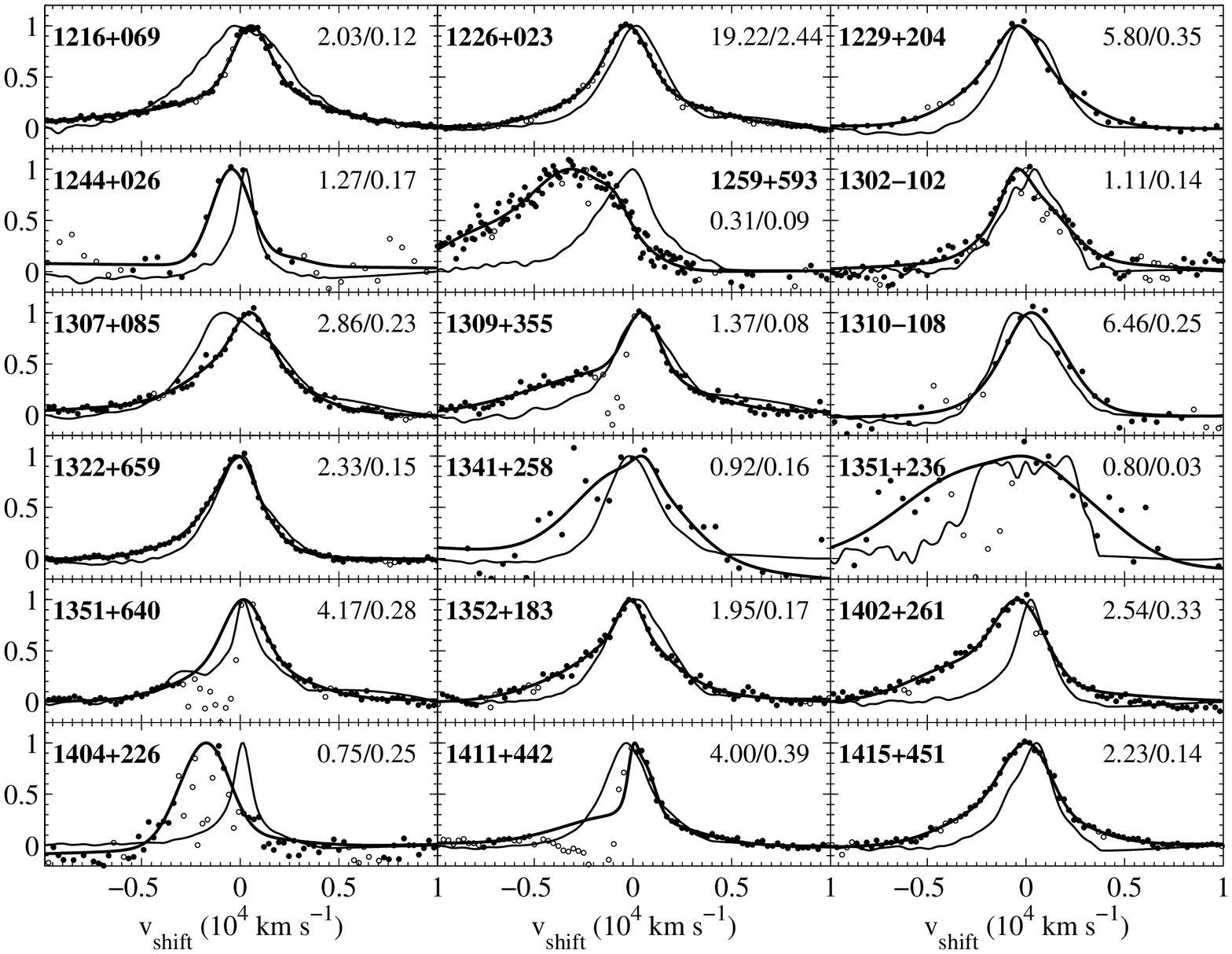}
\includegraphics[width=110mm,angle=-90]{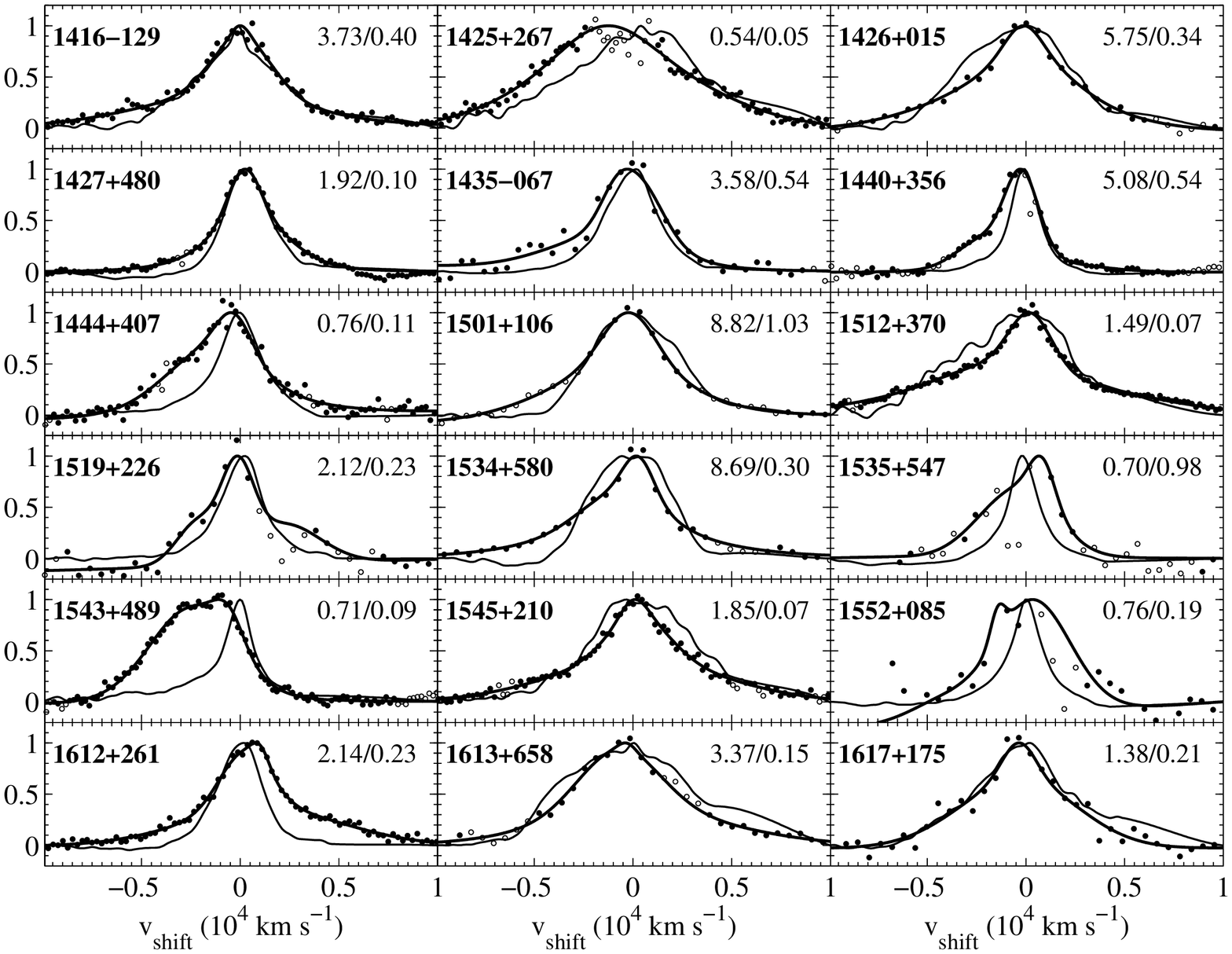}
\caption{--- Continued}
\end{figure*}
\setcounter{figure}{0}
\begin{figure*}
\includegraphics[width=62mm,angle=-90]{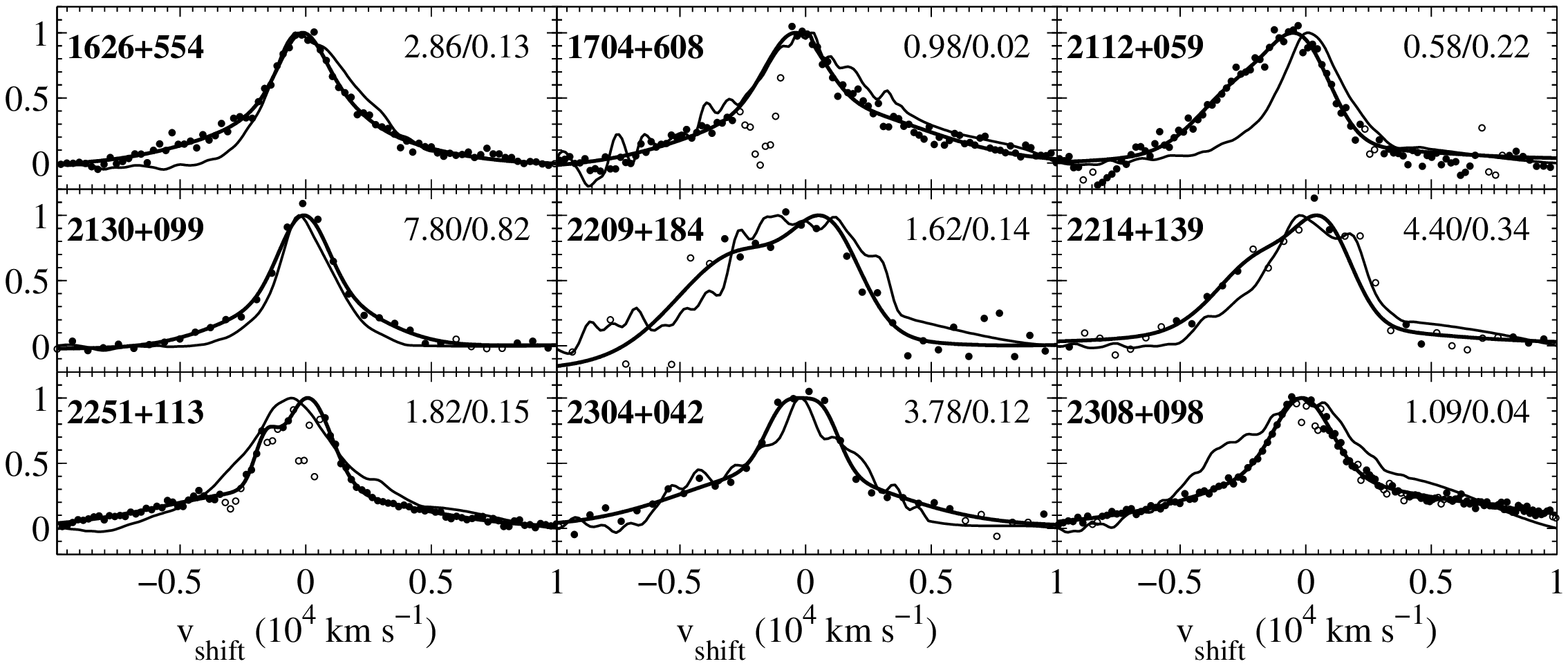}
\caption{--- Continued}
\end{figure*}

\subsection{The measured parameters}
The measured \C\ profile parameters are listed in Table 1,
together with some additional parameters. Specifically, column (2)
lists the redshifts determined using the peak of the \Ox\
$\lambda$5007 line. Column (3) lists the log of the black hole
mass \Mbh\ in units of $M_{\sun}$, estimated using $\nu
L_{\nu}$(3000\AA) and the \Hb\ FWHM (equation 3 in Laor 1998).
Column (4) lists the log of \Ledd\ (BL04). Column (5) lists the
\C\ EW in units of \AA\ (BL04). Column (6) lists the continuum
flux density at 1549\AA\ in units of
$10^{-15}$~erg~cm$^{-2}$~s$^{-1}$~\AA$^{-1}$. Column (7) lists the
\C\ FWHM in units of km~s$^{-1}$. Column (8) lists the shift of
the centroid of \C\ at 3/4 maximum from the rest wavelength in
units of the FWHM. This is the shift parameter defined by BG92.
Negative values indicate a blueshift. The rest wavelength of \C\
is taken as 1549.05~\AA, which is the mean doublet wavelength for
optically thin emission (2:1 flux ratio of the
1548.19~\AA/1550.77~\AA\ lines). Column (9) lists the shape
parameter for \C, defined by BG92 as (FW1/4M +
FW3/4M)/(2$\times$FWHM). Column (10) lists the line asymmetry
parameter, \C\ asymmetry, defined by De~Roberies (1985) as the
shift between the centroids at 1/4 and 3/4 maximum in units of the
FWHM. Positive values indicate excess blue wing flux. Column (11)
lists the \C/\Hb\ line flux ratio, R~$f_{\rm line}$ (throughout
this paper the prefix `R' indicates the ratio of a variable to
that of \Hb). The \Hb\ line flux was calculated using the \Hb\ EW
(BG92) and continuum flux density at 4861\AA\ measured here.
Column (12) lists the log of R~\C\ FWHM, calculated using the \Hb\
FWHM values from BG92 and the \C\ FWHM values in column (7).
Column (13) lists the ratio of the \C\ to \Hb\ line flux density
at $v_{\rm shift}=0$. Column (14) lists the continuum flux density
at 4861\AA\ in units of
$10^{-15}$~erg~cm$^{-2}$~s$^{-1}$~\AA$^{-1}$. Column (15) lists
the optical to UV slope \aouv, measured between 4861~\AA\ and
1549~\AA\ using $f_{\nu}$. Columns (16) and (17) list the EW of
the narrow, \Ox\ like, component of \C\ and \Hb, respectively, in
units of \AA. Finally, column (18) lists the correction factor for
the flux densities of the optical spectra (Section 2.2.1).

\begin{table*}
\begin{flushleft}
\begin{minipage}{180mm}
\caption{Emission-line and continuum properties.} \setlength{\tabcolsep}{0.7ex}
\begin{tabular}{@{}l*{18}{r}@{}}
\hline Object & \multicolumn{1}{c}{$z$}& \multicolumn{1}{c}{\Mbh}&
\multicolumn{1}{c}{\Ledd} & \multicolumn{9}{c}{\C} &
\multicolumn{1}{c}{$f_{\rm cont}$}&\multicolumn{1}{c}{$\alpha_{\rm
o,UV}$}& \multicolumn{2}{c}{N.C.
EW}  & \multicolumn{1}{c}{corre-} \\
& & & & \multicolumn{1}{c}{EW\fnrepeat{fn:1a}} &
\multicolumn{1}{c}{$f_{\rm cont}$} & \multicolumn{1}{c}{FWHM} &
\multicolumn{1}{c}{shift} & \multicolumn{1}{c}{shape} &
\multicolumn{1}{c}{asymm} &\multicolumn{1}{c}{R $f_{\rm
line}$}&\multicolumn{1}{c}{R FWHM} & \multicolumn{1}{c}{R $f$(0)}&
\multicolumn{1}{c}{(4861)}& &\multicolumn{1}{c}{\C} &
\multicolumn{1}{c}{\Hb} &\multicolumn{1}{c}{ction}\\
\multicolumn{1}{c}{(1)}
&\multicolumn{1}{c}{(2)}&\multicolumn{1}{c}{(3)}&\multicolumn{1}{c}{(4)}&
\multicolumn{1}{c}{(5)} &\multicolumn{1}{c}{(6)}
&\multicolumn{1}{c}{(7)} &\multicolumn{1}{c}{(8)}
&\multicolumn{1}{c}{(9)} &\multicolumn{1}{c}{(10)}&
\multicolumn{1}{c}{(11)}&
\multicolumn{1}{c}{(12)} &\multicolumn{1}{c}{(13)} &\multicolumn{1}{c}{(14)} &\multicolumn{1}{c}{(15)}
&\multicolumn{1}{c}{(16)} &\multicolumn{1}{c}{(17)}&\multicolumn{1}{c}{(18)} \\
\hline
0003+158 & $ 0.4505 $ & $  9.055 $ & $ -0.358 $ & $  63.5 $ & $    9.7 $ & $  3490 $ & $-$0.007 & $  1.255 $ & $-$0.067 & $    4.85 $ & $ -0.135 $& $   20.79 $ & $   1.39 $ & $  -0.30 $ & $  1.3 $ & $  1.8 $ & $  1.28 $\\
0003+199 & $ 0.0260 $ & $  7.220 $ & $ -0.342 $ & $  60.1 $ & $   87.9 $ & $  2090 $ & $  0.049 $ & $  1.181 $ & $  0.202 $ & $    5.21 $ & $  0.106 $& $   17.64 $ & $  10.66 $ & $  -0.16 $ & $  1.4 $ & $  4.7 $ & $  1.20 $\\
0007+106 & $ 0.0893 $ & $  8.561 $ & $ -0.972 $ & $  59 $ & $   26.2 $ & $  3550 $ & $-$0.004 & $  1.151 $ & $  0.053 $ & $    4.54 $ & $ -0.158 $& $   16.69 $ & $   3.38 $ & $  -0.21 $ & --- & $  4.1 $ & $  1.42 $\\
0026+129 & $ 0.1452 $ & $  7.833 $ & $  0.053 $ & $  19.3 $ & $   35.9 $ & $  1600 $ & $  0.075 $ & $  1.674 $ & $  0.151 $ & $    3.41 $ & $ -0.067 $& $   12.64 $ & $   3.12 $ & $   0.14 $ & $  0.3 $ & $  2.9 $ & $  2.01 $\\
0043+039 & $ 0.3859 $ & $  8.952 $ & $ -0.648 $ & $   5.4 $ & $    2.1 $ & $  4490 $ & $-$0.142 & $  1.380 $ & $-$0.411 & $    0.14 $ & $ -0.072 $& $    0.57 $ & $   0.84 $ & $  -1.22 $ & --- & --- & $  1.29 $\\
0049+171 & $ 0.0643 $ & $  8.146 $ & $ -1.437 $ & $ 203 $ & $    7.7 $ & $  4980 $ & $  0.021 $ & $  1.053 $ & $-$0.025 & $   12.49 $ & $ -0.023 $& $   37.49 $ & $   0.92 $ & $  -0.14 $ & $  0.4 $ & $ 13.7 $ & $  1.80 $\\
0050+124 & $ 0.0587 $ & $  7.238 $ & $  0.162 $ & $  29.9 $ & $   25.5 $ & $  4430 $ & $-$0.040 & $  0.864 $ & $  0.381 $ & $    2.45 $ & $  0.553 $& $    6.39 $ & $   6.09 $ & $  -0.75 $ & $  1.9 $ & $  7.0 $ & $  1.45 $\\
0052+251 & $ 0.1544 $ & $  8.745 $ & $ -0.822 $ & $ 119.0 $ & $   14.6 $ & $  5610 $ & $-$0.019 & $  1.168 $ & $  0.024 $ & $    7.28 $ & $  0.033 $& $   17.62 $ & $   2.75 $ & $  -0.54 $ & $  0.5 $ & $  3.7 $ & $  1.59 $\\
0157+001 & $ 0.1632 $ & $  8.006 $ & $ -0.261 $ & $  43 $ & $   12.7 $ & $  4430 $ & $-$0.344 & $  1.090 $ & $  0.105 $ & $    2.74 $ & $  0.255 $& $    6.08 $ & $   2.54 $ & $  -0.59 $ & $  2.1 $ & $  8.6 $ & $  2.07 $\\
0804+761 & $ 0.1005 $ & $  8.352 $ & $ -0.300 $ & $  45 $ & $   48.6 $ & $  4120 $ & $-$0.051 & $  1.146 $ & $  0.067 $ & $    4.09 $ & $  0.128 $& $   10.91 $ & $   4.51 $ & $   0.08 $ & $  0.8 $ & $  3.6 $ & $  1.44 $\\
0838+770 & $ 0.1318 $ & $  7.992 $ & $ -0.493 $ & $  50 $ & $    8.1 $ & $  4370 $ & $  0.045 $ & $  1.083 $ & $  0.140 $ & $    2.79 $ & $  0.195 $& $    6.93 $ & $   1.43 $ & $  -0.48 $ & $  0.7 $ & $  3.3 $ & $  2.43 $\\
0844+349 & $ 0.0644 $ & $  7.759 $ & $ -0.479 $ & $  28 $ & $   31.0 $ & $  3870 $ & $  0.013 $ & $  1.034 $ & $  0.006 $ & $    2.91 $ & $  0.204 $& $    8.45 $ & $   3.97 $ & $  -0.20 $ & $  0.7 $ & $  2.1 $ & $  2.12 $\\
0921+525 & $ 0.0352 $ & $  7.206 $ & $ -0.802 $ & $ 186 $ & $   15.7 $ & $  2940 $ & $  0.166 $ & $  1.095 $ & $  0.021 $ & $    7.91 $ & $  0.142 $& $   24.06 $ & $   2.54 $ & $  -0.41 $ & $  2.9 $ & $ 10.4 $ & $  2.47 $\\
0923+129 & $ 0.0287 $ & $  7.233 $ & $ -0.665 $ & $  93 $ & $   22.6 $ & $  3900 $ & $  0.112 $ & $  1.086 $ & $  0.055 $ & $    6.90 $ & $  0.292 $& $   17.53 $ & $   4.03 $ & $  -0.49 $ & $  1.0 $ & $  6.4 $ & $  2.88 $\\
0923+201 & $ 0.1929 $ & $  9.094 $ & $ -1.134 $ & $  28 $ & $   17.3 $ & $  6200 $ & $  0.128 $ & $  1.159 $ & $-$0.172 & $    2.87 $ & $ -0.089 $& $   17.29 $ & $   1.06 $ & $   0.44 $ & $  1.1 $ & $  0.9 $ & $  0.89 $\\
0947+396 & $ 0.2059 $ & $  8.530 $ & $ -0.909 $ & $  55 $ & $   10.2 $ & $  3650 $ & $-$0.073 & $  1.204 $ & $-$0.020 & $    5.98 $ & $ -0.121 $& $   26.82 $ & $   0.85 $ & $   0.18 $ & $  0.8 $ & $  1.8 $ & $  1.18 $\\
0953+414 & $ 0.2341 $ & $  8.488 $ & $ -0.196 $ & $  54.9 $ & $   14.9 $ & $  3250 $ & $-$0.039 & $  1.304 $ & $-$0.074 & $    2.36 $ & $  0.016 $& $   10.67 $ & $   2.22 $ & $  -0.34 $ & $  1.4 $ & $  4.3 $ & $  1.55 $\\
1001+054 & $ 0.1610 $ & $  7.645 $ & $ -0.020 $ & $  34.9 $ & $    4.3 $ & $  2920 $ & $-$0.200 & $  1.074 $ & $  0.069 $ & $    1.37 $ & $  0.225 $& $    3.20 $ & $   1.22 $ & $  -0.90 $ & $  0.9 $ & $  2.5 $ & $  2.53 $\\
1011$-$040 & $ 0.0584 $ & $  7.190 $ & $ -0.146 $ & $  25 $ & $   19.4 $ & $  2960 $ & $-$0.114 & $  1.170 $ & $-$0.032 & $    3.76 $ & $  0.313 $& $    7.18 $ & $   2.86 $ & $  -0.33 $ & $  0.3 $ & $  2.9 $ & $  2.64 $\\
1012+008 & $ 0.1865 $ & $  8.069 $ & $ -0.320 $ & $  23 $ & $    5.8 $ & $  6500 $ & $-$0.076 & $  1.018 $ & $  0.031 $ & $    0.82 $ & $  0.391 $& $    2.11 $ & $   1.57 $ & $  -0.86 $ & $  2.9 $ & $  5.8 $ & $  2.00 $\\
1022+519 & $ 0.0449 $ & $  6.940 $ & $ -0.600 $ & $  38 $ & $   13.7 $ & $  3550 $ & $-$0.011 & $  1.245 $ & $  0.212 $ & $    4.99 $ & $  0.341 $& $    9.64 $ & $   1.52 $ & $  -0.08 $ & $  1.1 $ & $  3.4 $ & $  1.00 $\\
1048$-$090 & $ 0.3461 $ & $  9.022 $ & $ -0.679 $ & $  91 $ & $    4.0 $ & $  5190 $ & $  0.081 $ & $  1.367 $ & $-$0.046 & $    4.15 $ & $ -0.034 $& $   16.37 $ & $   1.07 $ & $  -0.86 $ & --- & $  1.6 $ & $  0.87 $\\
1048+342 & $ 0.1667 $ & $  8.241 $ & $ -0.687 $ & $  46 $ & $    2.4 $ & $  4060 $ & $-$0.141 & $  1.076 $ & $-$0.159 & $    0.38 $ & $  0.053 $& $    3.10 $ & $   1.25 $ & $  -1.43 $ & $  1.9 $ & $  2.4 $ & $  4.06 $\\
1049$-$006 & $ 0.3596 $ & $  8.989 $ & $ -0.630 $ & $  67.0 $ & $    6.7 $ & $  3660 $ & $-$0.119 & $  1.136 $ & $  0.051 $ & $    4.23 $ & $ -0.165 $& $   23.84 $ & $   0.99 $ & $  -0.34 $ & $  0.1 $ & $  4.7 $ & $  3.25 $\\
1100+772 & $ 0.3115 $ & $  9.112 $ & $ -0.749 $ & $  84.0 $ & $    7.0 $ & $  3610 $ & $  0.019 $ & $  1.451 $ & $  0.106 $ & $    4.06 $ & $ -0.232 $& $   19.58 $ & $   1.61 $ & $  -0.71 $ & $  0.8 $ & $  3.8 $ & $  1.75 $\\
1103$-$006 & $ 0.4232 $ & $  9.132 $ & $ -0.737 $ & $  37.2 $ & $    3.3 $ & $  4370 $ & $  0.120 $ & $  1.087 $ & $  0.104 $ & $    2.54 $ & $ -0.151 $& $   10.25 $ & $   0.60 $ & $  -0.52 $ & $  1.1 $ & $  2.7 $ & $  1.61 $\\
1114+445 & $ 0.1438 $ & $  8.415 $ & $ -0.927 $ & $  55.0 $ & $    4.9 $ & $  3830 $ & $  0.129 $ & $  1.216 $ & $  0.084 $ & $    1.46 $ & $ -0.077 $& $    5.96 $ & $   1.84 $ & $  -1.15 $ & --- & $  2.0 $ & $  2.18 $\\
1115+407 & $ 0.1542 $ & $  7.505 $ & $ -0.139 $ & $  25.9 $ & $   11.1 $ & $  3600 $ & $-$0.185 & $  1.122 $ & $  0.041 $ & $    4.54 $ & $  0.320 $& $    7.49 $ & $   1.04 $ & $   0.07 $ & --- & $  5.1 $ & $  2.02 $\\
1116+215 & $ 0.1765 $ & $  8.425 $ & $ -0.139 $ & $  40.5 $ & $   35.3 $ & $  4280 $ & $-$0.108 & $  1.213 $ & $  0.085 $ & $    2.01 $ & $  0.166 $& $    7.10 $ & $   4.07 $ & $  -0.11 $ & --- & $  4.5 $ & $  1.35 $\\
1119+120 & $ 0.0500 $ & $  7.280 $ & $ -0.462 $ & $  29 $ & $   26.2 $ & $  3660 $ & $-$0.057 & $  1.286 $ & $-$0.011 & $    4.55 $ & $  0.304 $& $    7.33 $ & $   3.36 $ & $  -0.20 $ & $  0.4 $ & $  2.6 $ & $  1.22 $\\
1121+422 & $ 0.2248 $ & $  7.856 $ & $ -0.232 $ & $  41.7 $ & $   15.9 $ & $  2000 $ & $-$0.046 & $  1.160 $ & $-$0.035 & $   10.47 $ & $ -0.046 $& $   43.18 $ & $   0.71 $ & $   0.71 $ & --- & $  3.3 $ & $  0.62 $\\
1126$-$041 & $ 0.0601 $ & $  7.598 $ & $ -0.434 $ & $  30 $ & $   23.3 $ & $  3860 $ & $  0.037 $ & $  1.000 $ & $-$0.242 & $    2.01 $ & $  0.254 $& $    5.52 $ & $   3.45 $ & $  -0.33 $ & $  1.1 $ & $  3.2 $ & $  2.26 $\\
1149$-$110 & $ 0.0489 $ & $  7.729 $ & $ -0.916 $ & $  82 $ & $   11.1 $ & $  4920 $ & $-$0.123 & $  1.073 $ & $  0.058 $ & $    3.65 $ & $  0.206 $& $    8.67 $ & $   2.12 $ & $  -0.55 $ & $  0.8 $ & $  4.1 $ & $  1.21 $\\
1151+117 & $ 0.1759 $ & $  8.435 $ & $ -0.801 $ & $  26.6 $ & $   12.8 $ & $  3170 $ & $  0.064 $ & $  1.472 $ & $-$0.205 & $    2.98 $ & $ -0.133 $& $   17.06 $ & $   0.90 $ & $   0.33 $ & $  1.1 $ & $  1.3 $ & $  0.75 $\\
1202+281 & $ 0.1654 $ & $  8.462 $ & $ -1.053 $ & $ 290.0 $ & $    2.0 $ & $  3190 $ & $-$0.216 & $  1.156 $ & $  0.092 $ & $    4.49 $ & $ -0.200 $& $   26.51 $ & $   0.90 $ & $  -1.31 $ & $  4.6 $ & $  5.3 $ & $  0.98 $\\
1211+143 & $ 0.0810 $ & $  7.831 $ & $  0.051 $ & $  55.7 $ & $   29.4 $ & $  1690 $ & $  0.012 $ & $  1.224 $ & $  0.068 $ & $    2.35 $ & $ -0.043 $& $    8.96 $ & $   8.32 $ & $  -0.90 $ & $  0.9 $ & $  3.1 $ & $  2.21 $\\
1216+069 & $ 0.3318 $ & $  8.954 $ & $ -0.609 $ & $  64.5 $ & $    8.9 $ & $  3120 $ & $  0.180 $ & $  1.354 $ & $  0.033 $ & $    3.69 $ & $ -0.221 $& $   14.46 $ & $   1.78 $ & $  -0.60 $ & $  1.1 $ & $  1.4 $ & $  0.58 $\\
1226+023 & $ 0.1575 $ & $  8.876 $ & $ -0.012 $ & $  23.0 $ & $  211.0 $ & $  3430 $ & $-$0.085 & $  1.222 $ & $-$0.004 & $    1.51 $ & $ -0.011 $& $    7.40 $ & $  28.41 $ & $  -0.25 $ & --- & $  1.2 $ & $  2.69 $\\
1229+204 & $ 0.0640 $ & $  8.004 $ & $ -0.804 $ & $  48 $ & $   29.9 $ & $  4010 $ & $-$0.103 & $  1.134 $ & $  0.005 $ & $    3.99 $ & $  0.077 $& $   14.86 $ & $   3.40 $ & $  -0.10 $ & --- & $  1.8 $ & $  1.15 $\\
1244+026 & $ 0.0480 $ & $  6.614 $ & $  0.235 $ & $  17 $ & $   10.1 $ & $  2410 $ & $-$0.175 & $  1.034 $ & $-$0.024 & $    2.27 $ & $  0.463 $& $    7.81 $ & $   1.82 $ & $  -0.50 $ & $  1.1 $ & $  7.8 $ & $  2.45 $\\
1259+593 & $ 0.4770 $ & $  8.738 $ & $ -0.085 $ & $  15.3 $ & $    7.9 $ & $  6920 $ & $-$0.437 & $  1.071 $ & $  0.215 $ & $    1.97 $ & $  0.310 $& $    1.58 $ & $   1.00 $ & $  -0.19 $ & --- & --- & $  1.58 $\\
1302$-$102 & $ 0.2783 $ & $  8.749 $ & $ -0.080 $ & $  13.1 $ & $   20.8 $ & $  3580 $ & $-$0.061 & $  1.077 $ & $-$0.101 & $    2.85 $ & $  0.022 $& $    9.18 $ & $   3.41 $ & $  -0.42 $ & $  0.2 $ & $  2.2 $ & $  3.43 $\\
1307+085 & $ 0.1545 $ & $  8.541 $ & $ -0.651 $ & $  71.2 $ & $   11.0 $ & $  3680 $ & $  0.129 $ & $  1.190 $ & $  0.119 $ & $    2.71 $ & $ -0.057 $& $   11.66 $ & $   2.26 $ & $  -0.62 $ & $  0.5 $ & $  3.0 $ & $  1.02 $\\
1309+355 & $ 0.1825 $ & $  8.155 $ & $ -0.421 $ & $  33.5 $ & $   11.3 $ & $  3320 $ & $  0.117 $ & $  1.484 $ & $  0.421 $ & $    4.49 $ & $  0.053 $& $   15.41 $ & $   1.65 $ & $  -0.32 $ & $  0.6 $ & $  2.5 $ & $  1.39 $\\
1310$-$108 & $ 0.0343 $ & $  7.759 $ & $ -1.183 $ & $  78 $ & $   18.8 $ & $  3790 $ & $  0.092 $ & $  1.051 $ & $-$0.011 & $    5.66 $ & $  0.019 $& $   23.00 $ & $   2.63 $ & $  -0.28 $ & $  1.2 $ & $  7.0 $ & $  1.70 $\\
1322+659 & $ 0.1676 $ & $  8.076 $ & $ -0.409 $ & $  52.6 $ & $    8.5 $ & $  2990 $ & $-$0.055 & $  1.103 $ & $  0.089 $ & $    4.27 $ & $  0.030 $& $   15.00 $ & $   1.36 $ & $  -0.40 $ & --- & $  1.0 $ & $  1.73 $\\
1341+258 & $ 0.0864 $ & $  7.878 $ & $ -0.756 $ & $  62 $ & $    4.8 $ & $  5460 $ & $-$0.017 & $  1.040 $ & $  0.024 $ & $    1.82 $ & $  0.254 $& $    5.53 $ & $   1.37 $ & $  -0.90 $ & $  2.1 $ & $  1.8 $ & $  1.06 $\\
1351+236 & $ 0.0553 $ & $  8.216 $ & $ -1.748 $ & $ 101 $ & $    4.5 $ & $ 10250 $ & $-$0.105 & $  1.004 $ & $  0.028 $ & $    8.80 $ & $  0.195 $& $   22.92 $ & $   2.28 $ & $  -1.40 $ & --- & $  2.5 $ & $  1.80 $\\
1351+640 & $ 0.0880 $ & $  8.656 $ & $ -1.058 $ & $  43.3 $ & $   20.4 $ & $  3310 $ & $  0.052 $ & $  1.098 $ & $  0.004 $ & $    3.20 $ & $ -0.233 $& $   12.67 $ & $   4.76 $ & $  -0.73 $ & --- & $  3.5 $ & $  1.86 $\\
1352+183 & $ 0.1508 $ & $  8.299 $ & $ -0.629 $ & $  45.1 $ & $   10.3 $ & $  3350 $ & $-$0.049 & $  1.252 $ & $  0.131 $ & $    2.71 $ & $ -0.031 $& $   10.21 $ & $   1.29 $ & $  -0.18 $ & --- & $  1.4 $ & $  2.23 $\\
1402+261 & $ 0.1643 $ & $  7.845 $ & $  0.018 $ & $  30.3 $ & $   25.0 $ & $  4090 $ & $-$0.121 & $  1.232 $ & $  0.282 $ & $    6.31 $ & $  0.330 $& $    7.76 $ & $   1.58 $ & $   0.41 $ & --- & --- & $  3.23 $\\
1404+226 & $ 0.0978 $ & $  6.713 $ & $  0.232 $ & $  23.3 $ & $    6.2 $ & $  3260 $ & $-$0.538 & $  0.991 $ & $-$0.002 & $    2.35 $ & $  0.569 $& $    1.19 $ & $   1.13 $ & $  -0.51 $ & --- & --- & $  0.99 $\\
1411+442 & $ 0.0897 $ & $  7.874 $ & $ -0.535 $ & $  56.9 $ & $   11.5 $ & $  3080 $ & $  0.088 $ & $  0.814 $ & $  0.137 $ & $    2.32 $ & $  0.062 $& $   10.35 $ & $   2.66 $ & $  -0.72 $ & --- & $  2.1 $ & $  1.31 $\\
1415+451 & $ 0.1133 $ & $  7.797 $ & $ -0.579 $ & $  57.3 $ & $    9.5 $ & $  3730 $ & $-$0.016 & $  1.151 $ & $  0.118 $ & $    7.67 $ & $  0.154 $& $   18.63 $ & $   1.22 $ & $  -0.21 $ & --- & $  0.1 $ & $  1.33 $\\
1416$-$129 & $ 0.1292 $ & $  9.002 $ & $ -0.845 $ & $ 168.1 $ & $    6.2 $ & $  4090 $ & $  0.001 $ & $  1.139 $ & $  0.083 $ & $    1.65 $ & $ -0.174 $& $    7.65 $ & $   3.65 $ & $  -1.53 $ & --- & $  3.2 $ & $  6.04 $\\
1425+267 & $ 0.3635 $ & $  9.317 $ & $ -1.280 $ & $  64.8 $ & $    4.1 $ & $  8620 $ & $  0.161 $ & $  0.606 $ & $  0.276 $ & $    2.37 $ & $ -0.038 $& $    9.76 $ & $   1.20 $ & $  -0.93 $ & --- & $  3.5 $ & $  1.28 $\\
1426+015 & $ 0.0863 $ & $  8.921 $ & $ -1.117 $ & $  32 $ & $   60.7 $ & $  4910 $ & $-$0.021 & $  1.192 $ & $  0.070 $ & $    4.26 $ & $ -0.143 $& $   16.77 $ & $   5.23 $ & $   0.14 $ & $  0.4 $ & $  0.6 $ & $  1.71 $\\
1427+480 & $ 0.2203 $ & $  7.978 $ & $ -0.344 $ & $  53.2 $ & $    7.0 $ & $  2880 $ & $  0.094 $ & $  1.159 $ & $-$0.055 & $    4.69 $ & $  0.054 $& $   17.78 $ & $   0.58 $ & $   0.18 $ & $  1.1 $ & $  3.3 $ & $  1.51 $\\
1435$-$067 & $ 0.1288 $ & $  8.300 $ & $ -0.412 $ & $  39 $ & $   21.1 $ & $  3750 $ & $-$0.051 & $  1.125 $ & $  0.082 $ & $    1.77 $ & $  0.071 $& $    6.04 $ & $   3.30 $ & $  -0.38 $ & --- & $  1.3 $ & $  2.91 $\\
1440+356 & $ 0.0777 $ & $  7.335 $ & $ -0.013 $ & $  30.1 $ & $   29.4 $ & $  2530 $ & $-$0.125 & $  1.193 $ & $  0.211 $ & $    4.61 $ & $  0.241 $& $    8.91 $ & $   3.00 $ & $  -0.01 $ & $  0.6 $ & $  4.5 $ & $  1.35 $\\
1444+407 & $ 0.2676 $ & $  8.158 $ & $ -0.122 $ & $  17.9 $ & $   11.6 $ & $  4340 $ & $-$0.143 & $  1.081 $ & $  0.133 $ & $    3.03 $ & $  0.243 $& $    6.08 $ & $   0.95 $ & $   0.19 $ & --- & --- & $  0.84 $\\
1501+106 & $ 0.0365 $ & $  8.482 $ & $ -1.172 $ & $  64 $ & $   44.2 $ & $  4710 $ & $-$0.058 & $  1.163 $ & $  0.085 $ & $    2.12 $ & $ -0.065 $& $    8.12 $ & $   9.68 $ & $  -0.67 $ & $  0.2 $ & $  5.1 $ & $  2.20 $\\
\hline
\end{tabular}
\end{minipage}
\end{flushleft}
\end{table*}
\setcounter{table}{0}
\begin{table*}
\begin{flushleft}
\begin{minipage}{180mm}
\caption{--- Continued} \setlength{\tabcolsep}{0.7ex}
\begin{tabular}{@{}l*{18}{r}@{}}
\hline Object & \multicolumn{1}{c}{$z$}& \multicolumn{1}{c}{\Mbh}&
\multicolumn{1}{c}{\Ledd} & \multicolumn{9}{c}{\C} &
\multicolumn{1}{c}{$f_{\rm cont}$}&\multicolumn{1}{c}{$\alpha_{\rm
o,UV}$}& \multicolumn{2}{c}{N.C.
EW}  & \multicolumn{1}{c}{corre-} \\
& & & & \multicolumn{1}{c}{EW\footnote{EW values with a decimal
point are based on {\it HST} spectra, while the integer rounded
values are based on {\it IUE} spectra.\label{fn:1a}}} &
\multicolumn{1}{c}{$f_{\rm cont}$} & \multicolumn{1}{c}{FWHM} &
\multicolumn{1}{c}{shift} & \multicolumn{1}{c}{shape} &
\multicolumn{1}{c}{asymm} &\multicolumn{1}{c}{R $f_{\rm
line}$}&\multicolumn{1}{c}{R FWHM} & \multicolumn{1}{c}{R $f$(0)}&
\multicolumn{1}{c}{(4861)}& &\multicolumn{1}{c}{\C} &
\multicolumn{1}{c}{\Hb} &\multicolumn{1}{c}{ction}\\
\multicolumn{1}{c}{(1)}
&\multicolumn{1}{c}{(2)}&\multicolumn{1}{c}{(3)}&\multicolumn{1}{c}{(4)}&
\multicolumn{1}{c}{(5)} &\multicolumn{1}{c}{(6)}
&\multicolumn{1}{c}{(7)} &\multicolumn{1}{c}{(8)}
&\multicolumn{1}{c}{(9)} &\multicolumn{1}{c}{(10)}&
\multicolumn{1}{c}{(11)}& \multicolumn{1}{c}{(12)}
&\multicolumn{1}{c}{(13)} &\multicolumn{1}{c}{(14)}
&\multicolumn{1}{c}{(15)}
&\multicolumn{1}{c}{(16)} &\multicolumn{1}{c}{(17)}&\multicolumn{1}{c}{(18)} \\
\hline
1512+370 & $ 0.3713 $ & $  9.168 $ & $ -0.867 $ & $  84.3 $ & $    6.7 $ & $  4450 $ & $  0.021 $ & $  1.399 $ & $  0.186 $ & $    4.79 $ & $ -0.185 $& $   20.36 $ & $   0.96 $ & $  -0.31 $ & $  0.5 $ & $  5.9 $ & $  2.45 $\\
1519+226 & $ 0.1357 $ & $  7.777 $ & $ -0.311 $ & $  68 $ & $    7.3 $ & $  3080 $ & $-$0.052 & $  1.400 $ & $-$0.153 & $    4.54 $ & $  0.141 $& $    9.37 $ & $   1.05 $ & $  -0.30 $ & --- & $  5.6 $ & $  2.35 $\\
1534+580 & $ 0.0305 $ & $  8.047 $ & $ -1.565 $ & $  79 $ & $   29.7 $ & $  4010 $ & $  0.015 $ & $  1.141 $ & $  0.153 $ & $    6.12 $ & $ -0.125 $& $   24.89 $ & $   3.96 $ & $  -0.24 $ & --- & $  5.4 $ & $  1.37 $\\
1535+547 & $ 0.0389 $ & $  7.010 $ & $ -0.373 $ & $  27.6 $ & $    5.1 $ & $  3690 $ & $  0.132 $ & $  0.986 $ & $  0.233 $ & $    0.29 $ & $  0.396 $& $    0.59 $ & $   4.42 $ & $  -1.87 $ & $  1.6 $ & $  4.0 $ & $  1.72 $\\
1543+489 & $ 0.4009 $ & $  7.844 $ & $  0.369 $ & $  25.6 $ & $    8.7 $ & $  5330 $ & $-$0.364 & $  1.020 $ & $  0.079 $ & $    5.82 $ & $  0.533 $& $    5.68 $ & $   0.48 $ & $   0.53 $ & --- & --- & $  1.05 $\\
1545+210 & $ 0.2643 $ & $  9.165 $ & $ -0.925 $ & $  90.5 $ & $    6.7 $ & $  4530 $ & $  0.041 $ & $  1.212 $ & $-$0.007 & $    4.23 $ & $ -0.191 $& $   23.50 $ & $   1.48 $ & $  -0.69 $ & $  1.2 $ & $  2.7 $ & $  1.61 $\\
1552+085 & $ 0.1191 $ & $  7.364 $ & $  0.040 $ & $  47 $ & $    5.5 $ & $  4900 $ & $-$0.005 & $  1.130 $ & $  0.040 $ & $    3.51 $ & $  0.535 $& $    4.23 $ & $   1.59 $ & $  -0.91 $ & $  1.4 $ & $  2.7 $ & $  1.11 $\\
1612+261 & $ 0.1308 $ & $  7.913 $ & $ -0.395 $ & $  94.6 $ & $    6.5 $ & $  3810 $ & $  0.114 $ & $  1.266 $ & $-$0.118 & $    2.53 $ & $  0.180 $& $    7.60 $ & $   1.39 $ & $  -0.64 $ & --- & $ 12.9 $ & $  1.12 $\\
1613+658 & $ 0.1291 $ & $  8.953 $ & $ -1.457 $ & $  54 $ & $   22.6 $ & $  5730 $ & $-$0.104 & $  1.084 $ & $-$0.056 & $    4.76 $ & $ -0.168 $& $   19.69 $ & $   2.35 $ & $  -0.02 $ & $  0.1 $ & $  1.3 $ & $  1.14 $\\
1617+175 & $ 0.1137 $ & $  8.729 $ & $ -0.880 $ & $  34 $ & $   11.9 $ & $  4500 $ & $-$0.076 & $  1.155 $ & $  0.013 $ & $    1.48 $ & $ -0.074 $& $    6.60 $ & $   2.38 $ & $  -0.59 $ & --- & $  1.1 $ & $  2.37 $\\
1626+554 & $ 0.1317 $ & $  8.371 $ & $ -0.940 $ & $  45.6 $ & $   16.6 $ & $  3510 $ & $-$0.025 & $  1.234 $ & $  0.002 $ & $    4.65 $ & $ -0.107 $& $   22.12 $ & $   1.14 $ & $   0.34 $ & --- & $  0.5 $ & $  0.92 $\\
1704+608 & $ 0.3721 $ & $  9.198 $ & $ -0.772 $ & $  34.8 $ & $    9.4 $ & $  4340 $ & $-$0.086 & $  1.326 $ & $-$0.155 & $    8.25 $ & $ -0.180 $& $   46.54 $ & $   1.42 $ & $  -0.35 $ & $  0.3 $ & $  2.7 $ & $  1.16 $\\
2112+059 & $ 0.4597 $ & $  8.834 $ & $  0.116 $ & $  25.5 $ & $    6.4 $ & $  4790 $ & $-$0.159 & $  0.999 $ & $  0.141 $ & $    0.90 $ & $  0.176 $& $    2.33 $ & $   1.49 $ & $  -0.73 $ & --- & --- & $  1.74 $\\
2130+099 & $ 0.0631 $ & $  7.805 $ & $ -0.367 $ & $  47 $ & $   35.0 $ & $  3120 $ & $-$0.020 & $  1.154 $ & $-$0.007 & $    3.38 $ & $  0.127 $& $    9.47 $ & $   4.51 $ & $  -0.21 $ & $  0.2 $ & $  1.5 $ & $  1.07 $\\
2209+184 & $ 0.0697 $ & $  8.601 $ & $ -1.353 $ & $  54 $ & $   12.4 $ & $  7420 $ & $-$0.091 & $  0.970 $ & $  0.116 $ & $    2.77 $ & $  0.058 $& $   11.03 $ & $   2.11 $ & $  -0.45 $ & $  3.6 $ & $  2.7 $ & $  2.95 $\\
2214+139 & $ 0.0657 $ & $  8.308 $ & $ -1.027 $ & $  45 $ & $   28.1 $ & $  5200 $ & $-$0.001 & $  0.978 $ & $  0.150 $ & $    2.80 $ & $  0.058 $& $   11.80 $ & $   4.21 $ & $  -0.34 $ & $  2.1 $ & $  0.6 $ & $  0.87 $\\
2251+113 & $ 0.3255 $ & $  8.816 $ & $ -0.363 $ & $  66.0 $ & $    7.7 $ & $  3580 $ & $-$0.093 & $  1.209 $ & $  0.061 $ & $    3.49 $ & $ -0.065 $& $   13.12 $ & $   1.78 $ & $  -0.72 $ & --- & $  2.1 $ & $  0.93 $\\
2304+042 & $ 0.0426 $ & $  8.320 $ & $ -1.633 $ & $ 176 $ & $    6.8 $ & $  4040 $ & $-$0.044 & $  1.381 $ & $  0.146 $ & $    5.55 $ & $ -0.206 $& $   27.08 $ & $   2.19 $ & $  -1.01 $ & $  6.3 $ & $  0.6 $ & $  1.04 $\\
2308+098 & $ 0.4336 $ & $  9.372 $ & $ -0.936 $ & $  81.5 $ & $    5.1 $ & $  3970 $ & $-$0.036 & $  1.342 $ & $-$0.023 & $    5.26 $ & $ -0.303 $& $   23.22 $ & $   0.87 $ & $  -0.46 $ & --- & $  1.6 $ & $  1.61 $\\
\hline
\end{tabular}
\end{minipage}
\end{flushleft}
\end{table*}

\section{RESULTS \& DISCUSSION}

\subsection{The \C\ versus \Hb\ profile parameters}

Figure 2 compares the distribution of the EW and the profile
parameters: shape, asymmetry, and shift for \C\ and \Hb. The \C\
line shows a broader distribution in all parameters. In
particular, the \C\ EW (in \AA) ranges from 5 to 290, with a mean
and dispersion of $59\pm 45$, while \Hb\ EW ranges from 23 to 230
with a mean and dispersion of $99\pm 37$, i.e. the dispersion/mean
ratio in \C\ is twice as large as in \Hb. The asymmetry parameter
distribution of \Hb\ shows a small excess of negative values,
which correspond to red asymmetric profiles (i.e. excess flux on
the red wing), while \C\ shows a large excess of blue asymmetric
profiles. Similarly, the shift parameter in \C\ extends to a blue
shift of $-0.538$, while in \Hb\ the maximum blue shift is only
$-0.135$ (see similar result in fig. 3 of Sulentic, Marziani \&
Dultzin-Hacyan 2000). In both lines the maximum shift to the red
is small. This confirms the various earlier literature results on
the systematic differences between the high and low ionization
line profiles (Section 1).

\begin{figure}
\includegraphics[width=88mm]{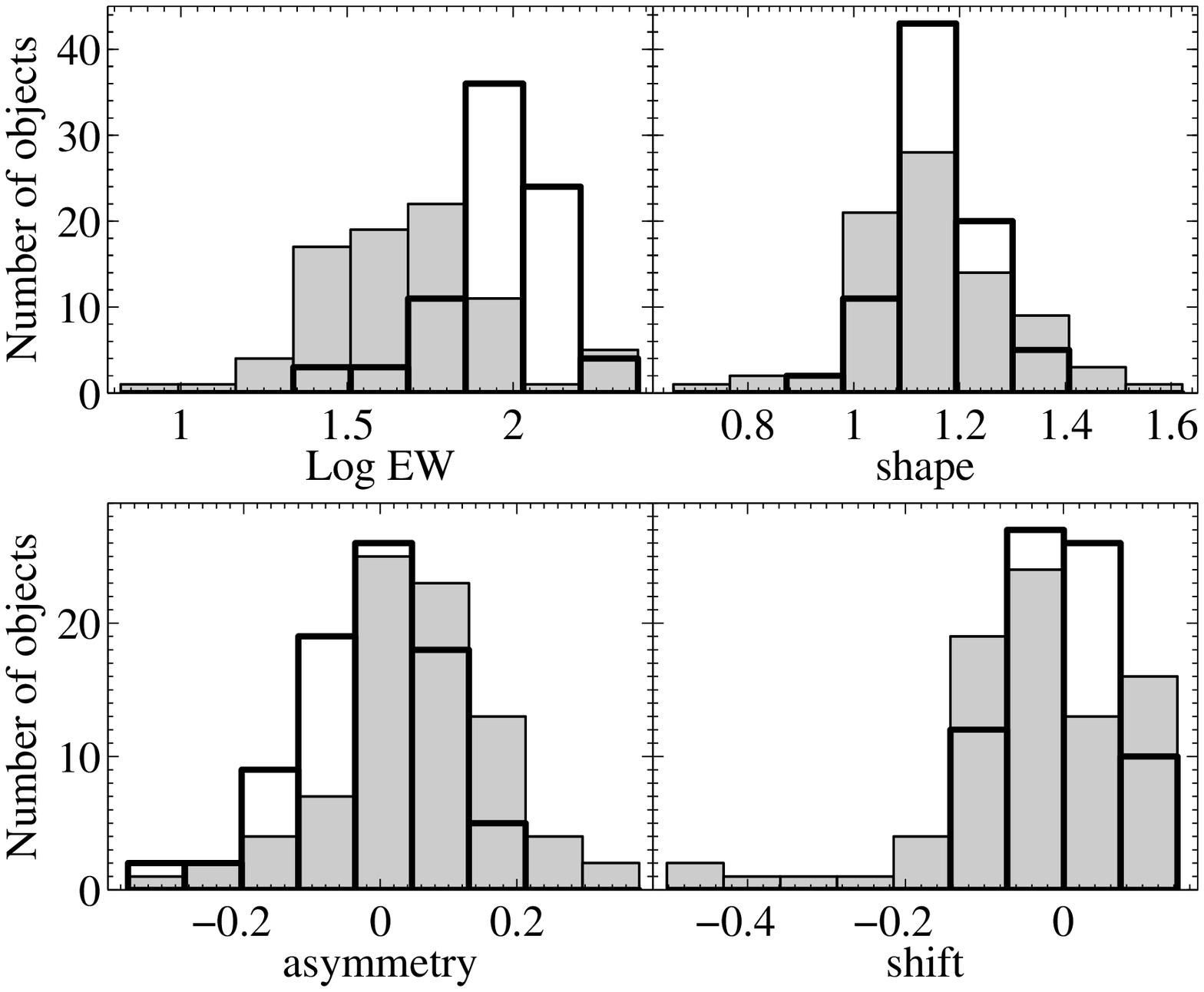}
\caption{The distributions of $\log$~EW, shape, asymmetry and
shift of the \Hb\ line (transparent) and  the \C\ line (gray)
profiles. Note the larger range in EW and shape parameter values
for \C, its tendency to show blue excess flux asymmetry (positive
asymmetry parameter), and the `tail' of objects with large shifts
to the blue (negative shift parameter).}
\end{figure}

\subsection{The \C\ versus \Hb\ FWHM}

Figure 3 presents the \C\ FWHM versus \Hb\ FWHM. The two lines
show quite different distributions of line widths. The \C\ line
shows a smaller relative range of FWHM (1600-10,250~km~s$^{-1}$,
a ratio of 6.4) compared to \Hb\ (830-9400~km~s$^{-1}$, ratio of
11.3), with most objects having FWHM in
the range of 3000-5000~km~s$^{-1}$, compared to a rather even
distribution in the range 1500-8000~km~s$^{-1}$ for \Hb.
Furthermore, there is a distinct lack of AGN with a narrow \C.
About 21 per cent (17/81) of our objects qualify as `narrow line'
AGN based on \Hb, i.e. have FWHM~$<$~2000~km~s$^{-1}$, but only
2.5 per cent (2/81) of the objects have a
\C~FWHM~$<$~2000~km~s$^{-1}$.

Figure 3 (upper panel) shows that although there is a significant
correlation between the \C\ and \Hb\ FWHM, as found in earlier
studies (Corbin 1991; Corbin \& Boroson 1996; Marziani et al.
1996), the scatter in the correlation is large. Most importantly,
the scatter is not random but rather shows a {\em systematic}
trend. Fig. 3 (lower panel) shows the ratio of \C/\Hb\ FWHM as a
function of \Hb\ FWHM. In most ($\sim 90$ per cent) objects with
\Hb\ FWHM~$<$~4000~km~s$^{-1}$ the \C\ line is {\em broader} than
\Hb, while in most objects where \Hb\ FWHM~$>$~4000~km~s$^{-1}$
the \C\ line is {\em narrower} than \Hb. A similar effect is seen
in fig.~7 of Shemmer et al. (2004), where they plot the
distribution of the \C\ FWHM versus \Hb\ FWHM for a sample 82
quasars, all at a relatively high redshift ($2<z<3.5$) compared to
our sample ($z<0.5$). This similarity is noteworthy given the
large difference in redshift and maximum luminosities of the two
samples (a similar effect is also seen in the low redshift sample
of Corbin \& Boroson 1996, fig.7 there).

A commonly accepted view is that the \C\ line tends to be
broader than \Hb\ (e.g. fig. 4 in Warner et al. 2003), which is
consistent with the smaller emission radius indicated in
some reverberation studies (see Section 3.3). However, this
appears not to hold in most of the \Hb\ FWHM~$>$~4000~km~s$^{-1}$
objects, where \C\ is narrower than \Hb.
A prominent example of such a reverese behaviour is found
in the subclass of double peaked Balmer line AGN, where \Hb\ is often
broader than \C\ (Halpern et al. 1996).

The systematic differences between the
\Hb\ and \C\ profiles raise the concern, can the
\C\ FWHM replace the \Hb\ FWHM as a comparable accuracy tool for estimating
\Mbh?

\begin{figure}
\includegraphics[width=88mm]{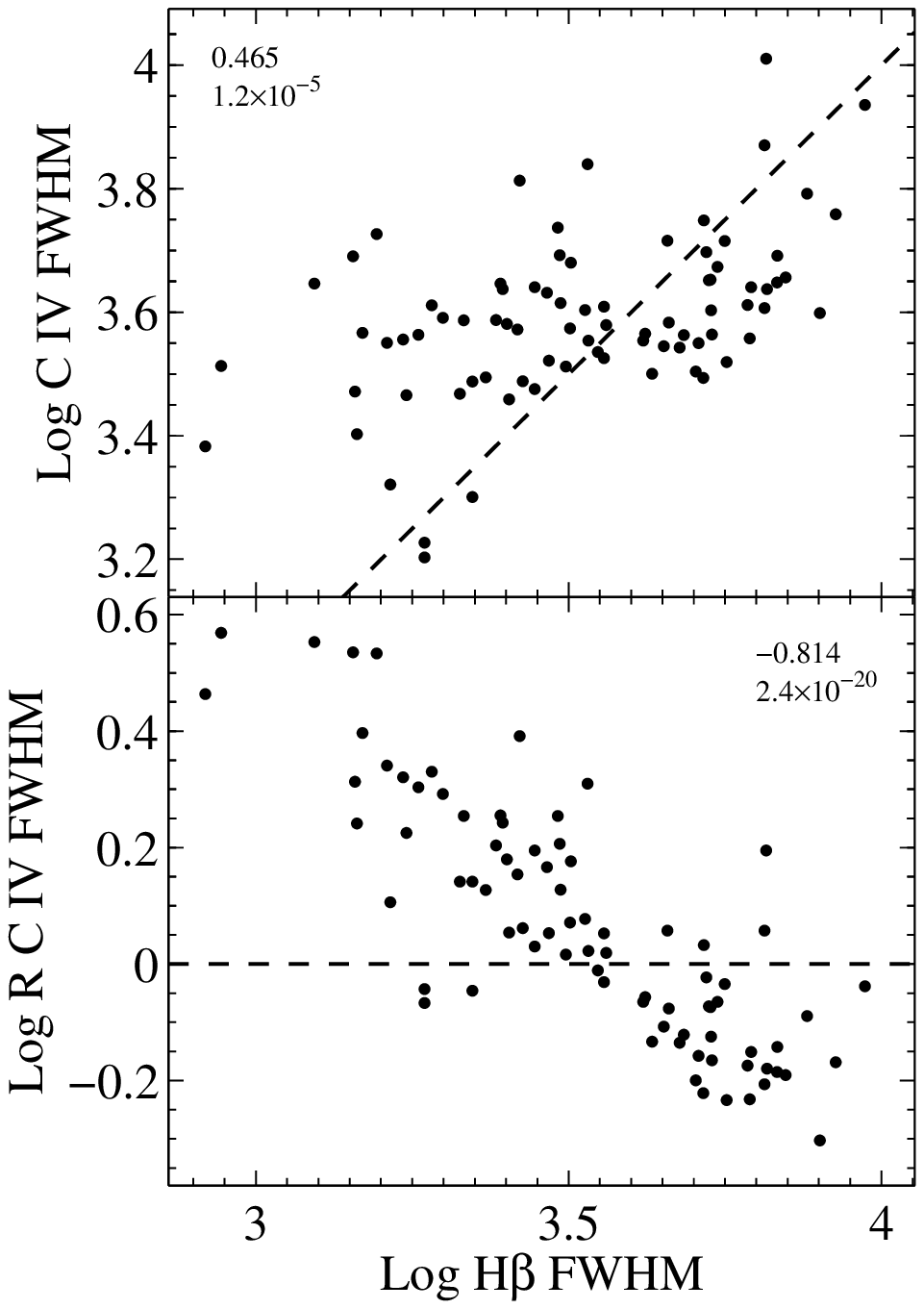}
\caption{Upper panel: the \C\ FWHM versus \Hb\ FWHM correlation.
The dashed line marks a 1:1 relation. Lower panel: the relation
between \C~FWHM/\Hb~FWHM ratio and the \Hb\ FWHM. The values of
\rS\ and Pr are indicated. Note that in nearly all objects where
\Hb\ FWHM~$<4000$~km~s$^{-1}$ \C\ is broader than \Hb, while in
most objects where \Hb\ FWHM~$>4000$~km~s$^{-1}$ \C\ is narrower
than \Hb.}
\end{figure}

\subsection{The \C\ based \Mbh\ estimates}

Current studies which utilize the \C\ FWHM to estimate \Mbh\
(Vestergaard 2002; Corbett et al. 2003;  Warner et~al. 2003, 2004;
Dietrich \& Hamann 2004) assume the same dependence of \Mbh\ on
luminosity and line width as used in the \Hb\ based \Mbh\
estimates (e.g. Kaspi et al. 2000). Intercalibration of the two
methods in the studies above yields a constant correction factor,
which corresponds to a mean \C\ emission radius in the BLR of
about half the \Hb\ emission radius, consistent with reverberation
results in some nearby AGN (Korista et al. 1995; Onken \& Peterson
2002). However, the systematic trend in the \C/\Hb\ FWHM ratio
shown in Fig. 3 indicates that the above \C\ based \Mbh\ estimates
will be biased. Since \Mbh~$\propto$~FWHM(line)$^2$, the \C\ based
\Mbh\ estimates in objects with  \Hb\ FWHM~$<$~4000~km~s$^{-1}$
will be too high by a factor of typically 3-4, and up to a factor
of 10 in the extreme cases. Similarly, in objects with \Hb\
FWHM~$>$~4000~km~s$^{-1}$ the above \C\ based \Mbh\ estimates will
be too low by the same factors (see Section 3.5).

As Fig.~3 (upper panel) shows, the \C\ FWHM is only weakly
dependent on the \Hb\ FWHM, which indicates that the above bias
cannot be reliably corrected based on the \C\ FWHM, e.g. an AGN
with \C\ FWHM~=~4000~km~s$^{-1}$ can have an \Hb\ FWHM anywhere
from ~1500~km~s$^{-1}$ to ~7000~km~s$^{-1}$ with about the same
probability. However, indirect correlations mentioned below
(Section 3.5) may allow somewhat more accurate estimates of the
likely \Hb\ FWHM for a given \C\ FWHM.

An additional, less direct, indication for the lower accuracy of
the \C\ \Mbh\ estimate is provided by the Baldwin relation. BL04
have shown that the Baldwin relation for \C\ appears to be induced
by a significantly tighter relation between the \C\ EW and \Ledd\
(\rS~$= -0.581$, Figure 4, left panel). We have repeated this
analysis with \Ledd\ estimated using the \C\ based expression for
\Mbh\ provided by Vestergaard (2002) \footnote{Note that here we
use $\nu L_{\nu}$(1549~\AA) instead of $\nu L_{\nu}$(1350~\AA)
used by Vestergaard, but the scatter between the two luminosities
is negligible.}. No significant correlation is present between the
\C\ EW and \Ledd\ derived using the Vestergaard relation (\rS~$=
-0.205$, Fig.~4, middle panel), which most likely results from the
large scatter present in the \C\ based \Mbh\ estimates. A
qualitatively similar result was found by Shemmer et al. (2004),
where they find a significant correlation between the BLR
metalicity and \Ledd\ in their $2<z<3.5$ sample, when \Mbh\ is
estimated using \Hb, but a significantly lower correlation when
\C\ is used to infer \Mbh\ (figs.~5 \& 8 there).

\begin{figure*}
\includegraphics[width=72mm,angle=-90]{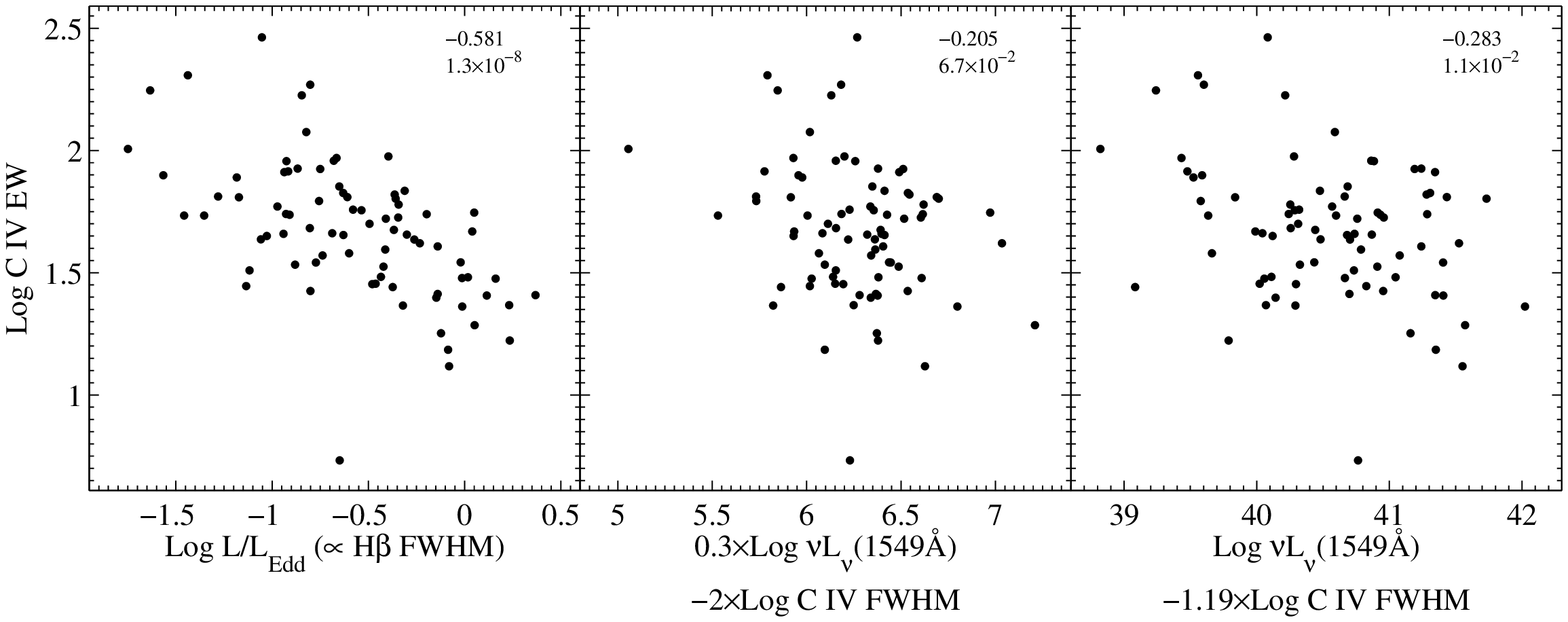}
\caption{An indirect indication for the accuracy of the \C\ based
\Mbh\ estimate (\rS\ and Pr are indicated in each panel). Left
panel: the correlation of the \C\ EW with \Ledd\ estimated based
on the \Hb\ FWHM and $\nu L_{\nu}$(3000~\AA) (taken from BL04).
Middle panel: the same correlation where \Ledd\ is estimated using
the Vestergaard (2002) relation for \Mbh, based on the \C\ FWHM
and the UV luminosity. Note the significant drop in \rS\ compared
to the \Hb\ based estimate.
 Right panel: the linear combination of log[$\nu L_{\nu}$(1549~\AA)]
and log(\C\ FWHM) which gives the largest \rS. Note that the
improved \rS\ is still not significant.}
\end{figure*}

The \C\ line may follow a different radius luminosity relation
than assumed by Vestergaard. To allow for that we looked for the
strongest correlation of log \C\ EW with a general linear
combination of the form log~[$\nu L_{\nu}$(1549~\AA)]~$+a\log$(\C\
FWHM), where $a$ is a free parameter. The strongest correlation we
find is \rS~$= -0.283$ at $a = -1.19$ (Fig.~4, right panel). This
correlation is still much weaker than the correlation with \Ledd\
estimated based on \Hb. This indicates that the scatter seen in
the \C\ versus \Hb\ FWHM relation (Fig.~3) is probably not just
due to a different radius versus luminosity power law relation for
the \C\ emitting region (e.g. if \Mbh~$\propto$~\C~FWHM$^2$ then
$a = -1.19$ implies $R_{\rm BLR}\propto L^{1.68}$ for \C, which
appears unlikely; the alternative option of maintaining $R_{\rm
BLR}\propto L^{0.5}$ for \C\ would require
\Mbh~$\propto$~\C~FWHM$^{0.6}$ which appears even less plausible).
{\em What is it then that determines the \C\ profile?}

\subsection{The correlation analysis}

To obtain clues for the effects which may control the \C\ profile,
and in particular the {\em differences} between the \C\ and \Hb\
profiles, we carried out an extensive correlation analysis, as
described below.  We use the Spearman rank-order correlation coefficient,
which tests for any monotonic relation, rather than the Pearson
correlation coefficient, which tests the significance of a
linear relation only.

Table 2 presents the correlation matrix. Rows are present for all
the items from Table 1 (except the narrow component EW of \C\ and \Hb),
all the optical emission line parameters from BG92 (see BG92 for
definitions), the values of
the optical to X-ray slope
(3000~\AA\ to 2~keV) \aox\ from Brandt, Laor \& Wills (2000) and
Laor \& Brandt (2002), $\nu L_{\nu}$(3000~\AA), \Mbh\ and \Ledd\
(see above), $\alpha_{\rm x}$, measured at 0.2-2~keV (Wang et al. 1996;
Laor et al. 1997b), the optical continuum slope $\alpha_{\rm o}$ from
Neugebauer et al. (1987, table 5 there), the degree of broadband
optical continuum polarization (3200-8600~\AA; Berriman et al.
1990, table 1 there), and the \C\ absorption EW (Laor \& Brandt 2002,
table 1 there). Columns are present for all the relevant
parameters measured
in this study, excluding the \C\ asymmetry, R asymmetry,
and R shift, which do not show any significant correlations.

Correlations where the corresponding null probability is less
than $10^{-4}$ are marked in bold face (excluding the \C\ absorption EW
correlations, where the two strongest correlations are marked). To
help interpret the
significance of the $r_S$ values listed in Table 2, we list
in Table 3 the null probability corresponding to various $r_S$ values and
sample sizes.

The \C\ EW shows the largest number of significant correlations.
These were discussed in BL04 and will not be repeated here. The
next \C\ profile parameter with the largest number of significant
correlations is the \C\ to \Hb\ FWHM ratio, which is further
discussed below.

\begin{table*}
\begin{minipage}{160mm}
\caption{Correlation Matrix.\fnrepeat{fn:2a}}
\setlength{\tabcolsep}{0.8ex}
\begin{tabular}{@{}l*{11}{r}@{}}
\hline & \multicolumn{10}{c}{\C} &  \multicolumn{1}{c}{$\alpha_{\rm o,UV}$}\\
    &   \multicolumn{1}{c}{EW} & \multicolumn{1}{c}{$L_{\rm line}$} &   \multicolumn{1}{c}{FWHM} &   \multicolumn{1}{c}{shift}
     &   \multicolumn{1}{c}{shape}    &   \multicolumn{1}{c}{R EW}    &   \multicolumn{1}{c}{R FWHM}
        &   \multicolumn{1}{c}{R shape}  &    \multicolumn{1}{c}{R $f_{\rm line}$}
            &   \multicolumn{1}{c}{R $f_{\rm line}$(0)}  \\
    \hline
\C~EW    &       &   0.096   &   0.079   &   0.376   &   0.159   &   {\bf 0.768}   &   {\bf $\bmath{-}$0.424}  &   0.148   &   0.402   &   {\bf 0.586}   &   $-$0.316  \\
$L_{\rm line}$   &   0.096   &       &   0.058   &   0.016   &   0.383   &   0.006   &   $\bmath{-0.503}$  &   0.311   &   0.104   &   0.276   &   0.194   \\
\C~FWHM &   0.079   &   0.058  &       &   $-$0.110  &   $-$0.342  &   $-$0.007  &   0.051   &   $-$0.280  &   $-$0.100  &   $-$0.145  &   $-$0.188  \\
\C~shift    &   0.376   &   0.016   &   $-$0.110  &       &   0.200   &   0.286   &   $-$0.310  &   0.257   &   0.094   &   0.321   &   $-$0.084  \\
\C~shape    &   0.159   &   0.383   &   $-$0.342  &   0.200   &       &   0.136   &   {\bf $\bmath{-}$0.487}  &   {\bf 0.863}   &   0.271   &   0.345   &   0.209   \\
R \C~EW    &   {\bf 0.768}   &  0.006    &   $-$0.007  &   0.286   &   0.136   &       &    $-$0.281   &   0.119   &   {\bf 0.580}   &   {\bf 0.577}   &   $-$0.334  \\
R \C~FWHM   &   {\bf $\bmath{-}$0.424}  &   $\bmath{-0.503}$  &   0.051   &   $-$0.310  &   {\bf $-$0.487}  &   $-$0.281  &     &   {\bf $-$0.446}  &   $-$0.169  &   {\bf $-$0.631}  &   0.065   \\
R \C~shape  &   0.148   &   0.311   &   $-$0.280  &   0.257   &   {\bf 0.863}   &   0.119   &   {\bf $\bmath{-}$0.446}  &       &   0.326   &   0.371   &   0.307   \\
R $f_{\rm line}$    &   0.402   &   0.104   &   $-$0.100  &   0.094   &   0.271   &   {\bf 0.580}   &   $-$0.169  &   0.326   &       &   {\bf 0.767}   &   {\bf 0.469}   \\
R $f_{\rm line}(0)$   &   {\bf 0.586}   &   0.276   &   $-$0.145  &   0.321   &   0.345   &   {\bf 0.577}   &   {\bf $\bmath{-}$0.631}  &   0.371   &   {\bf 0.767}   &       &   0.269   \\
\aouv   &   $-$0.316  &   0.194   &   $-$0.188  &   $-$0.084  &   0.209   &   $-$0.334  &   0.065   &   0.307   &   {\bf 0.469}   &   0.269   &       \\
$\nu L_{\nu}$(3000\AA)  &   $-$0.154  &   {\bf 0.883}  &   0.115   &   $-$0.084  &   0.333   &   $-$0.199  &   $-$0.418  &   0.221   &   $-$0.225  &   $-$0.040  &   $-$0.006  \\
\Ledd   &   {\bf $\bmath{-}$0.581}  &   0.088  &   $-$0.387  &   $-$0.330  &   $-$0.027  &   $-$0.391  &   $-$0.099\fnrepeat{fn:2b}   &   $-$0.116  &   $-$0.262  &   {\bf $\bmath{-}$0.555}  &   0.145   \\
\Mbh    &   0.215   &   0.234\fnrepeat{fn:2c}  &   0.382   &   0.083   &   0.289   &   0.070   &   $-$0.183\fnrepeat{fn:2b}  &   0.245   &   $-$0.067  &   0.316   &   $-$0.133  \\
$\rm M_{\rm V}$ &   0.179   &   $\bmath{-0.873}$   &   $-$0.126  &   0.089   &   $-$0.297  &   0.173   &   0.384   &   $-$0.196  &   0.206   &   0.043   &   0.030   \\
R   &   0.271   &   0.360   &   0.023   &   0.114   &   0.234   &   0.401   &   $-$0.314  &   0.186   &   0.088   &   0.210   &   $-$0.246  \\
\aox    &   {\bf 0.525}   &   0.146   &   $-$0.088  &   0.141   &   0.362   &   0.391   &   $-$0.367  &   0.406   &   {\bf 0.560}   &   {\bf 0.617}   &   0.249   \\
\Ox~EW  &   {\bf 0.624}   &   $-$0.026   &   0.009   &   0.373   &   0.062   &   {\bf 0.502}   &   $-$0.340  &   0.106   &   0.176   &   0.414   &   $-$0.265  \\
\Fe~EW  &   {\bf $\bmath{-}$0.518}  &   $-$0.249  &   0.046   &   $-$0.402  &   $-$0.301  &   {\bf $\bmath{-}$0.645}  &   {\bf 0.554}   &   $-$0.355  &   $-$0.385  &   {\bf $\bmath{-}$0.590}  &   0.218   \\
\Hb~FWHM    &   {\bf 0.427}   &  {\bf 0.427}   &   {\bf 0.465}   &   0.179   &   0.189   &   0.246   &   {\bf $\bmath{-}$0.814}  &   0.213   &   0.090   &   {\bf 0.496}   &   $-$0.171  \\
R \Ox\ peak &   {\bf 0.624}   &   0.140   &   0.078   &   0.367   &   0.180   &   {\bf 0.582}   &   {\bf $\bmath{-}$0.551}  &   0.225   &   0.352   &   {\bf 0.626}   &   $-$0.198  \\
\Hb~EW  &   0.347   &   0.132   &   0.097   &   0.172   &   0.043   &   $-$0.250  &   $-$0.289  &   0.017   &   $-$0.257  &   0.074   &   $-$0.009  \\
\He~EW  &   0.363   &   $-$0.315   &   $-$0.333  &   0.235   &   0.027   &   0.147   &   $-$0.011  &   0.060   &   0.092   &   0.180   &   $-$0.059  \\
$\rm M_{\Ox}$   &   $-$0.261  &   $\bmath{-0.707}$  &   $-$0.034  &   $-$0.237  &   $-$0.325  &   $-$0.139  &   {\bf 0.620}   &   $-$0.268  &   0.137   &   $-$0.256  &   0.195   \\
R \Fe    &   {\bf $\bmath{-}$0.626}  &   $-$0.381  &   $-$0.012  &   {\bf $\bmath{-}$0.453}  &   $-$0.349  &   {\bf $\bmath{-}$0.430}  &   {\bf 0.700}   &   $-$0.361  &   $-$0.198  &   {\bf $\bmath{-}$0.556}  &   0.195   \\
R \He    &   0.275   &   $-$0.365   &   $-$0.375  &   0.236   &   0.048   &   0.203   &   0.048   &   0.078   &   0.152   &   0.164   &   $-$0.033  \\
R \Ox    &   {\bf 0.471}   &   $-$0.117   &   $-$0.044  &   0.302   &   0.024   &   {\bf 0.595}   &   $-$0.218  &   0.096   &   0.282   &   0.410   &   $-$0.260  \\
\Hb\ shift\fnrepeat{fn:2d}   &   $-$0.053  &   0.145  &   $-$0.131  &   0.173   &   0.103   &   0.088   &   0.078   &   0.102   &   0.094   &   0.020   &   0.093   \\
\Hb\ shape   &   $-$0.184  &   $-$0.034  &   0.038   &   $-$0.146  &   $-$0.173  &   $-$0.084  &   0.120   &   {\bf $\bmath{-}$0.585}  &   $-$0.281  &   $-$0.249  &   $-$0.319  \\
\Hb\ asymmetry   &   $-$0.371  &   $-$0.397 &   0.076   &   $-$0.219  &   $-$0.276  &   $-$0.302  &   {\bf 0.475}   &   $-$0.185  &   $-$0.253  &   {\bf $\bmath{-}$0.475}  &   $-$0.038  \\
$\alpha_x$\fnrepeat{fn:2e}  &   0.263   &   0.280   &   0.360   &   0.337   &   0.050   &   0.145   &   {\bf $\bmath{-}$0.572}  &   0.059   &   $-$0.091  &   0.261   &   $-$0.215  \\
$\alpha_o$\fnrepeat{fn:2f}  &   $-$0.174  &   {\bf 0.597}  &   0.125   &   $-$0.158  &   0.161   &   $-$0.293  &   $-$0.124  &   0.068   &   $-$0.282  &   $-$0.082  &   0.006   \\
Pol. per cent &   0.324   &   0.063  &   0.275   &   0.271   &   $-$0.093  &   0.264   &   $-$0.317  &   $-$0.155  &   $-$0.160  &   0.068   &   {\bf $\bmath{-}$0.492}  \\
\C\ abs. EW\fnrepeat{fn:2g} &   $-$0.257  &   $-$0.322  &   0.115   &   0.002   &   $-$0.269  &   $-$0.142  &   0.162   &   $-$0.290  &  {\bf $\bmath{-}$0.380}  &   $-$0.342  &   {\bf $\bmath{-}$0.400}  \\
\hline
\end{tabular}
\footnotetext[1]{Correlations with null probability $<1\times
10^{-4}$ are in bold face (excluding the \C\ abs. EW correlations,
where the two strongest correlations are marked).\label{fn:2a}}
\footnotetext[2]{Correlation coefficient when \Hb~FWHM kept
fixed.\label{fn:2b}} \footnotetext[3]{Correlation coefficient when
$\nu L_{\nu}$(3000\AA) kept fixed.\label{fn:2c}}
\footnotetext[4]{75 objects were used in the
analysis.\label{fn:2d}} \footnotetext[5]{65 objects were used in
the analysis.\label{fn:2e}} \footnotetext[6]{69 objects were used
in the analysis.\label{fn:2f}} \footnotetext[7]{54 objects were
used in the analysis.\label{fn:2g}}
\end{minipage}
\end{table*}


\subsection{The R \C\ FWHM}

Figure 5 presents most of the significant correlations between R
\C\ FWHM and the various other parameters. The strongest
correlations are with parameters related to the Fe~II strength,
the \Ox\ strength, and the \Hb\ asymmetry. These parameters are
part of the BG92 set of correlated optical emission line
properties, which they termed eigenvector 1 (EV1). Narrow Line
Seyfert 1 galaxies, NLS1s, generally populate the extreme end of
EV1 (Osterbrock \& Pogge 1985, and citations thereafter), with
strong Fe~II emission, weak \Ox\ emission, and blue excess \Hb\
asymmetry. As shown in Fig. 5, in such objects the \C\ line is
broader than \Hb. Large R \C\ FWHM objects also have a steep
$\alpha_{\rm x}$, consistent with the strong correlation between
$\alpha_{\rm x}$ and EV1 (Laor et al. 1997b), and they also show
weaker \C\ emission, consistent with the known tendency of NLS1s
to show weaker \C\ emission (Wills et al. 1999; Shang et al.
2003). A new result shown in Fig. 5 is that the weakness of \C\
relative to \Hb\ is more pronounced when looking at the relative
line flux densities at $v=0$~km~s$^{-1}$. This probably results
from the combined effect of the broadening and weakening of the
\C\ emission in NLS1s. In addition, we find here that large R \C\
FWHM objects tend to have line shape parameters closer to unity.

\begin{figure*}
\includegraphics[width=49.5mm,angle=-90]{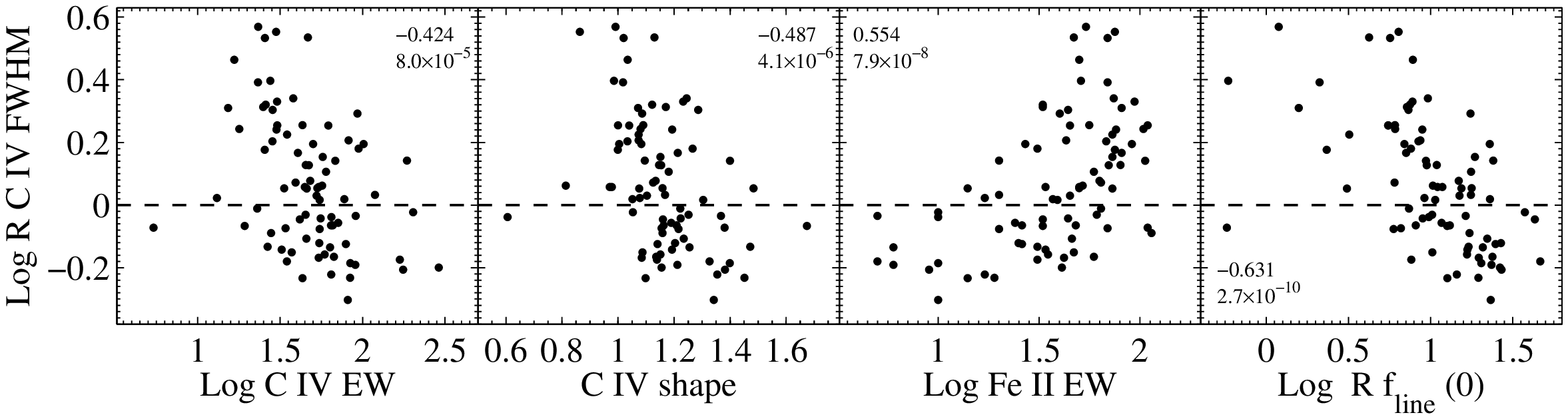}
\includegraphics[width=49.5mm,angle=-90]{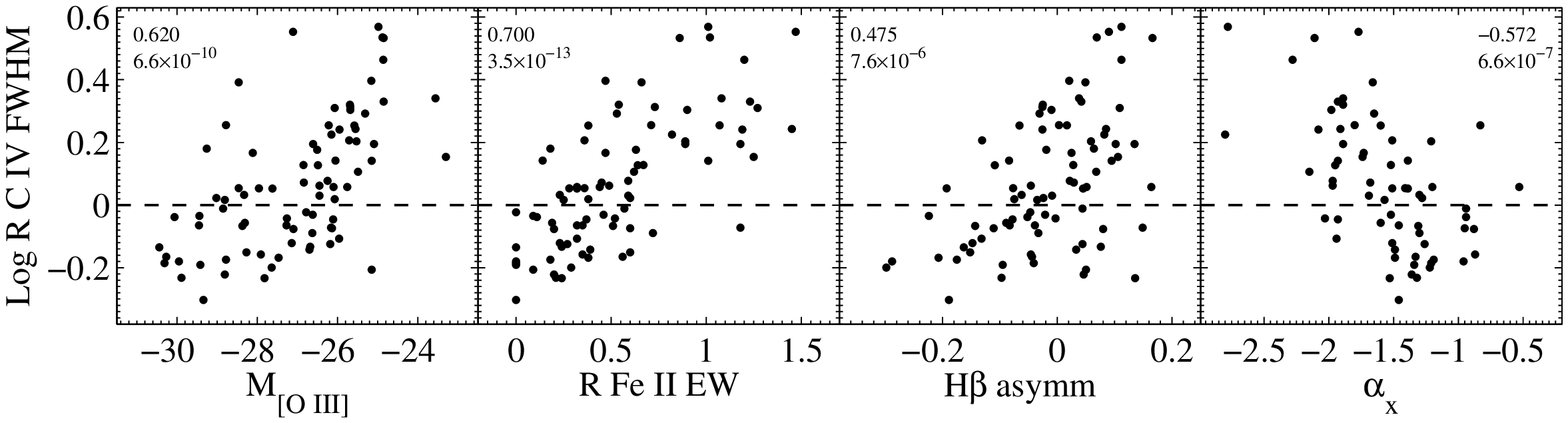}
\caption{The dependence of the \C\ to \Hb~FWHM ratio on various
parameters. The values of \rS\ and Pr are indicated in each panel.
Objects where \C\ is significantly broader than \Hb\ (above the
dashed line) tend to have weak \C, strong \Fe, weak \Ox, steep
$\alpha_{\rm X}$, and blue asymmetric \Hb, parameters which are
all related to the BG92 EV1.}
\end{figure*}

\begin{table}
\begin{minipage}{82mm}
\caption{Conversion of \rS\fnrepeat{fn:3a} to Pr\fnrepeat{fn:3b},
as a function of $N$\fnrepeat{fn:3c}.}
\begin{tabular}{@{}l*{5}{r}@{}}
\hline $|$\rS$|$ & \multicolumn{5}{c}{$N=$}\\
&   \multicolumn{1}{c}{81}  &   \multicolumn{1}{c}{75}  &   \multicolumn{1}{c}{69}  &   \multicolumn{1}{c}{65}  &   \multicolumn{1}{c}{54}  \\
\hline
0.1 &$  -0.427  $&$ -0.405  $&$ -0.383  $&$ -0.369  $&$ -0.326  $\\
0.2 &$  -1.134  $&$ -1.069  $&$ -1.003  $&$ -0.958  $&$ -0.833  $\\
0.3 &$  -2.187  $&$ -2.049  $&$ -1.911  $&$ -1.819  $&$ -1.560  $\\
0.4 &$  -3.666  $&$ -3.423  $&$ -3.18   $&$ -3.017  $&$ -2.564  $\\
0.45    &$  -4.602  $&$ -4.292  $&$ -3.98   $&$ -3.771  $&$ -3.194  $\\
0.5 &$  -5.697  $&$ -5.306  $&$ -4.914  $&$ -4.652  $&$ -3.927  $\\
0.55    &$  -6.98   $&$ -6.495  $&$ -6.008  $&$ -5.682  $&$ -4.783  $\\
0.6 &$  -8.491  $&$ -7.894  $&$ -7.294  $&$ -6.894  $&$ -5.789  $\\
0.65    &$  -10.287 $&$ -9.555  $&$ -8.822  $&$ -8.332  $&$ -6.981  $\\
0.7 &$  -12.45  $&$ -11.556 $&$ -10.661 $&$ -10.063 $&$ -8.414  $\\
0.8 &$  -18.48  $&$ -17.132 $&$ -15.783 $&$ -14.883 $&$ -12.402 $\\
0.9 &$  -29.493 $&$ -27.312 $&$ -25.13  $&$ -23.675 $&$ -19.667 $\\
\hline
\end{tabular}
\footnotetext[1]{Spearman rank-order correlation
coefficient.\label{fn:3a}} \footnotetext[2]{Null probability,
marked here on log scale.\label{fn:3b}} \footnotetext[3]{Sample
size.\label{fn:3c}}
\end{minipage}
\end{table}

We note in passing that the correlations of the \C~EW, \C\
luminosity and \C\ shape with R \C\ FWHM  can be applied for
improving the accuracy of the \C\ based \Mbh\ estimates.
Specifically, the measured values of \C~EW, shape and luminosity,
can be used to calculate the likely value of R \C\ FWHM, e.g.
based on a linear regression fit, and that value can then be used
to infer the likely value of \Hb~FWHM for a given \C~FWHM. This
estimated \Hb~FWHM can then be used to calculate \Mbh, thus
overcoming part of the large scatter present in \C\ based \Mbh\
estimates. A search for the tightest correlation between
$\log$~\Hb~FWHM and a linear combination of  $\log$~\C~FWHM,
$\log$(\C~EW), $\log
L$(\C) and \C\ shape yields a best fit relation
\begin{equation}
\begin{array}{l}
\log({\rm H\beta~FWHM_{est}})= \log({\rm\C~FWHM})\\
+0.25\log({\rm\C~EW}) +0.13\log[L({\rm\C})]\\
+0.4({\rm\C~shape})-6.6,
\end{array}
\end{equation}
where the EW is in units of \AA, and $L({\rm\C})$ is in units of
erg~s$^{-1}$. Figure 6, upper panel, shows the \Mbh\ estimates
based on \C\ (following Vestergaard 2002) versus the estimate
based on \Hb, which shows a dispersion of  0.505 (log scale),
while the lower panel shows instead a comparison with \Mbh\
derived using ${\rm H\beta~FWHM_{est}}$ rather than \C~FWHM. The
dispersion reduces to 0.378, which is smaller, but still
considerable. Thus, in the absence of direct observations of \Hb,
the UV based ${\rm H\beta~FWHM_{est}}$ may provide a somewhat more
accurate estimate of \Mbh\ than provided by the \C~FWHM.

\begin{figure}
\includegraphics[width=80mm]{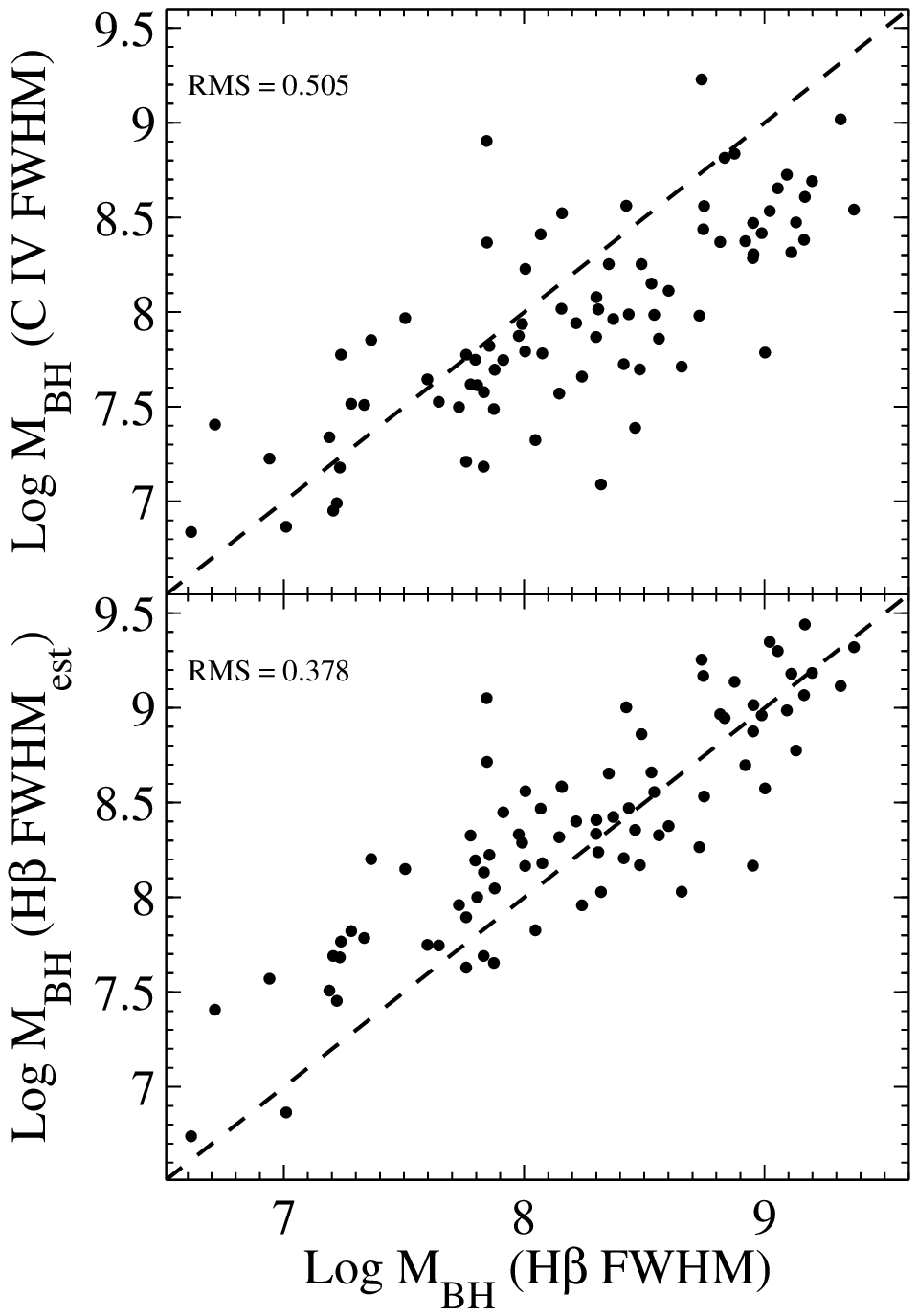}
\caption{Upper panel: the \Mbh\ estimates based on \C\ (following
Vestergaard 2002) versus the estimate based on \Hb. Note the large
dispersion (0.505 in log scale) measured around a 1:1 relation
(noted by the dashed line). Lower panel: as above, using the ${\rm
H\beta~FWHM_{est}}$ rather than the \C~FWHM. Note the reduced but
still significant dispersion (0.378). Thus, in the absence of
direct observations of \Hb, the UV based ${\rm H\beta~FWHM_{est}}$
may provide a somewhat more accurate estimate of \Mbh\ than
provided by the \C~FWHM.}
\end{figure}

The EV1 correlations are commonly interpreted as physical changes
in the AGN properties which are related to the value of \Ledd\
(e.g. BG92). Figure 7 shows there is indeed a strong correlation
of R \C\ FWHM with \Ledd\ (\rS\ = 0.604). A plausible physical
explanation for this effect is an outflowing wind component in the
BLR, driven by a large \Ledd, which can produce a higher column
density outflow and thus a higher line emissivity. Such an outflow
will have a lower density (due to expansion and increased speed),
and would contribute mostly to the higher ionization lines,
producing a broadened, shifted, and blue asymmetric \C\ line
(Section 1). The lower ionization lines, such as \Hb, may
originate from denser lower ionization gas, lying deeper within
the BLR structure, which would be less affected by radiation
pressure, and mostly dominated by gravity.
 However, Fig. 7 shows there is an even
stronger correlation of R \C\ FWHM with \Mbh\ (\rS~=~$-$0.770),
for which we find no clear physical explanation. Both correlations
shown in Fig. 7 could however be induced by the tight relation between R
\C\ FWHM and \Hb\ FWHM (\rS~$= -0.814$; Table 2), and the fact
that \Ledd\ and \Mbh\ are correlated with the \Hb\ FWHM. The
partial correlations of R \C\ FWHM with  \Ledd\ and \Mbh, for a
fixed \Hb\ FWHM, are indeed insignificant in our sample (\rS\ of
$-0.099$ and $-0.183$, respectively). This does not preclude the
above scenario of a physical relation between R \C\ FWHM and
\Ledd, but in order to test it one needs a sample of AGN covering
a wide range of \Ledd\ at a fixed \Hb\ FWHM. This will allow one
to separate out the dependence of R \C\ FWHM
on \Ledd\ (or \Mbh) from its
dependence on \Hb\ FWHM. To achieve that, one needs a sample of
AGN covering a wide range in luminosity (e.g. Yuan \& Wills 2003,
figs. 4 \& 5 there).

\begin{figure}
\includegraphics[width=46mm,angle=-90]{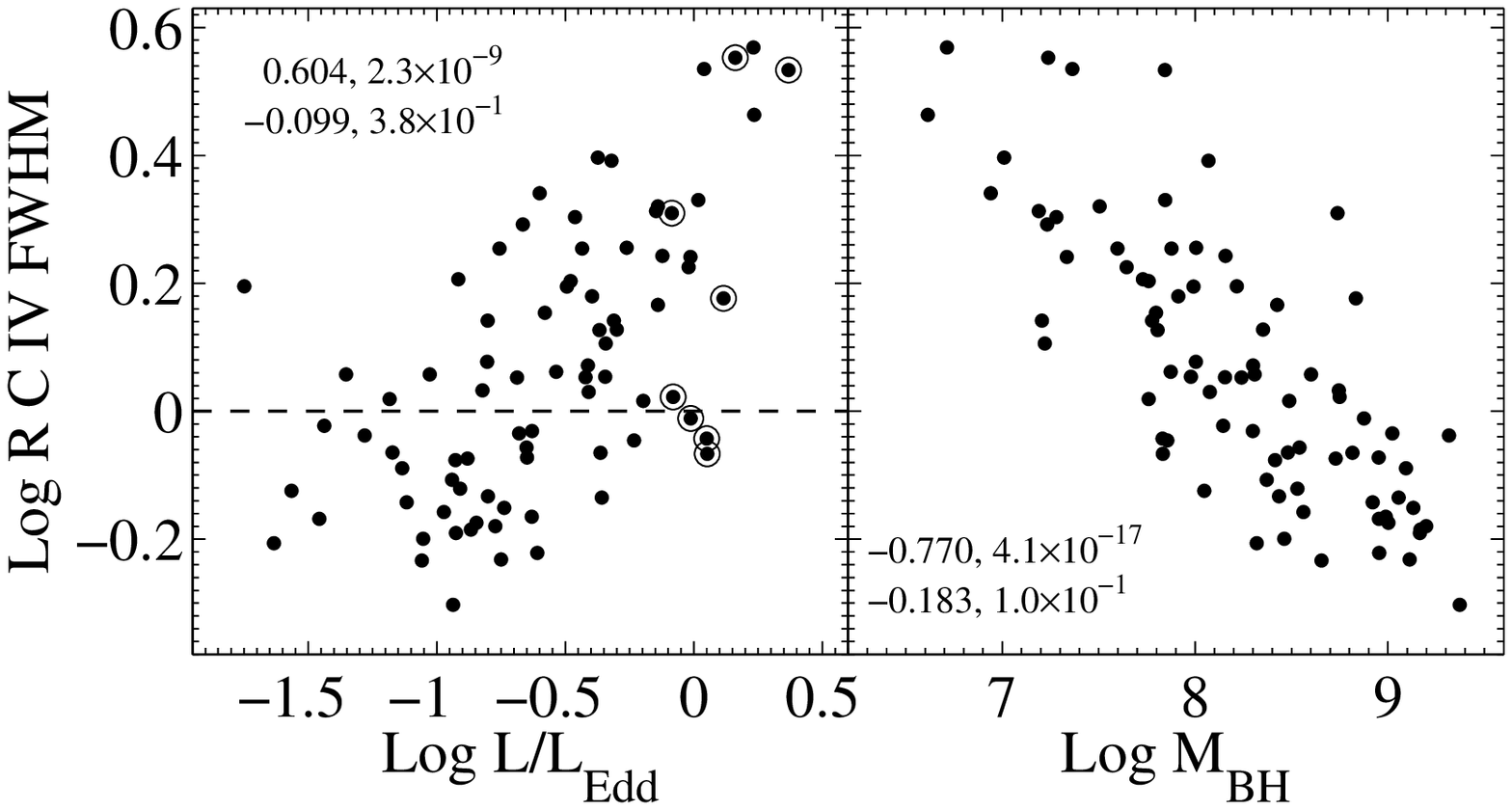}
\caption{Left panel: the relation between R \C~FWHM and \Ledd. A
high R \C~FWHM occurs only at a high \Ledd, but there are high
\Ledd\ objects where R \C~FWHM is low (the profiles of the circled
objects appear in Fig.~8).
 Right panel: the relation between R \C~FWHM and \Mbh.
The values of \rS\ and Pr are indicated in two rows for each plot.
The upper row is for the plotted correlation, and the lower row
for the partial correlation at a fixed \Hb~FWHM. Both partial
correlations are insignificant, indicating the apparently strong
correlations plotted here may be induced by the correlation of the
\Hb~FWHM with \Ledd, and \Mbh.}
\end{figure}

Despite the plausibility of the high \Ledd\ driven BLR outflow
scenario, Fig. 7 clearly demonstrates that although a high \Ledd\
is necessary for a large  R \C\ FWHM (there are no large R \C\
FWHM objects at low \Ledd), it is clearly not sufficient (there
are low R \C\ FWHM objects at a high \Ledd). This is further
demonstrated in Figure 8 which shows two different types of line
profiles in objects with a high \Ledd\ ($\ga 0.8$). The left panel
shows the \C\ and \Hb\ profiles of I~Zw~1 and three more
`I~Zw~1-like' objects in our sample. All objects have a large
values of R \C\ FWHM, a large \C\ blue-shift, and a large blue
excess asymmetry (see Leighly \& Moore 2004 for two additional
examples). Such objects are found only at high \Ledd\ values (Fig.
7), which supports the \Ledd\ driven outflow interpretation.
However, the right panel in Fig. 8 shows the profiles of four
other high \Ledd\ objects where R \C\ FWHM is small ($\la 0$), the
\C\ line shift is small, and the line profiles are symmetric.
Thus, having a high \Ledd\ appears to be a necessary, but not
sufficient, condition for having `I~Zw~1-like' \C\ profiles. What is
the difference between the `normal' and `I~Zw~1-like' objects at a
high \Ledd?  A correlation analysis of the various emission
parameters with R \C\ FWHM for the subset of \Ledd~$>0.5$ objects
did not yield new correlations beyond those present for the
complete sample. We note, anecdotally, that all the objects on the
left panel of Fig. 8 have relatively weak soft X-ray emission
(PG~1259+593, PG~1543+489, Brandt et~al. 2000), or show
significant X-ray and UV absorption (PG~0050+124, Gallo et al.
2004, Purquet et al. 2004; PG~2112+059, Gallagher et al. 2004a),
while none of the objects on the right panel show evidence for
either effects. This trend is consistent with the
result of Gallagher et al. (2004b) that AGN with a blueshifted
\C\ emission show excess X-ray absorption compared to normal AGN.
A possible interpretation of the above trend is
that the outflow component indicated by the `I~Zw~1-like' \C\
profile also tends to absorb the UV and/or X-ray emission.
However, it does not explain why radiation pressure does not
produce such an outflow in a significant fraction of high \Ledd\
AGN.

\begin{figure*}
\includegraphics[width=90mm,angle=-90]{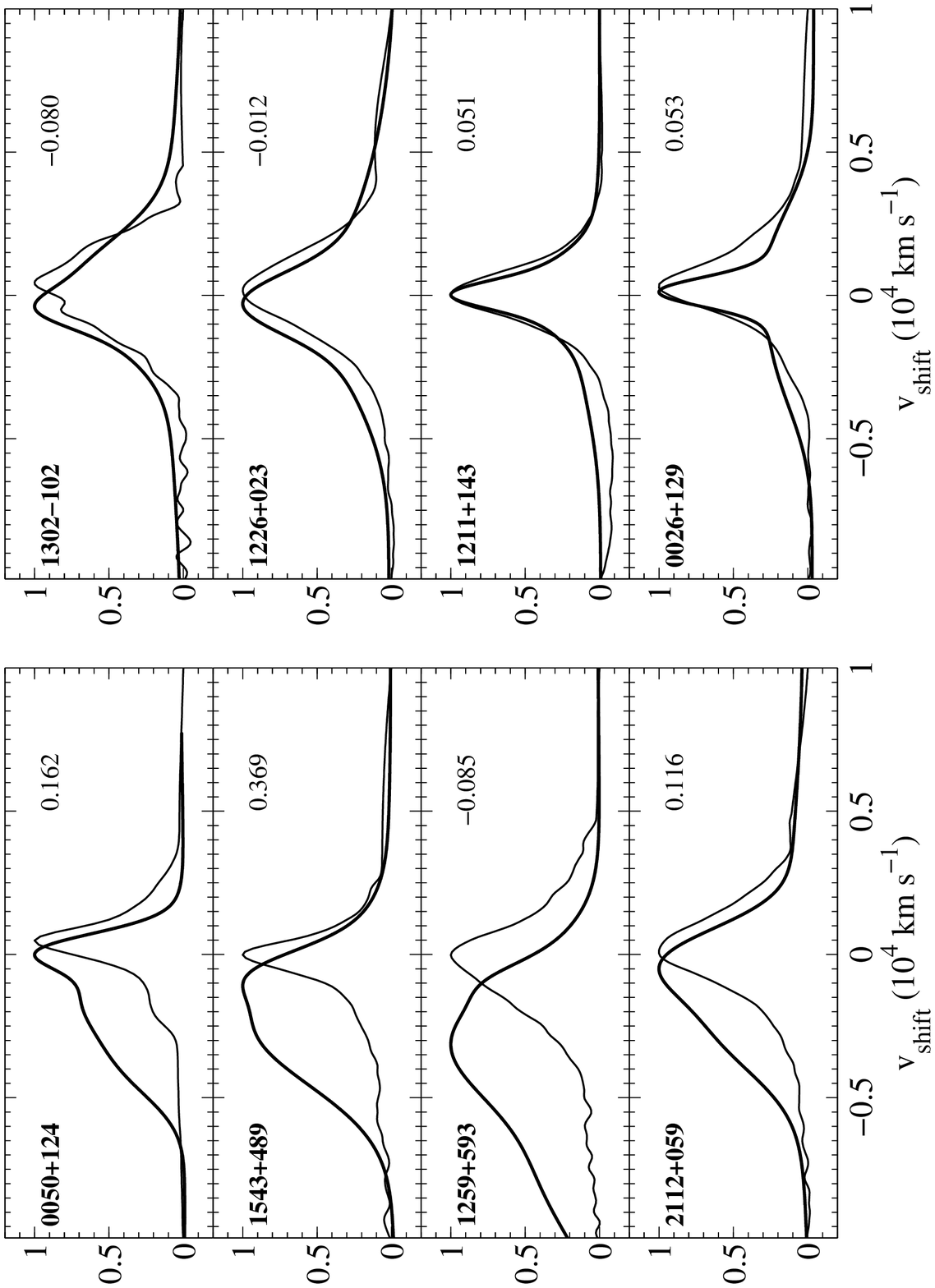}
\caption{Not all high \Ledd\ ($\ga 0.8$) objects are alike. Left
panel: four high \Ledd\ objects with `I Zw 1-like' \C\ profiles
(upper left object is I~Zw~1), i.e.~where \C\ is strongly
blueshifted and with a blue asymmetric profile. A similar, but
significantly weaker effect is seen in \Hb.  Right panel: four
high \Ledd\ objects where the \C\ and \Hb\ profiles are quite
similar. The name of the object is indicated at the left side of
each panel, and log~\Ledd\ is indicated at the right side of each
panel.}
\end{figure*}

\subsection{Other correlations}

Below we discuss various other correlations found in our sample,
and compare them to some correlations found in earlier studies.

\subsubsection{\C\ Shift versus EW}

Espey et al. (1989) noted an inverse relation between the shift of
the \C\ line peak, measured with respect to H$\rm\alpha$, and the
\C\ EW, such that the line peak becomes more blueshifted as it
becomes weaker. Similar correlations were found by Corbin (1990),
where the \C\ shift was measured with respect to \Mg, and by
Marziani et al.\footnote{Note that Marziani et al. separate the
\C\ profile into narrow and broad components based on a Gaussian
decomposition, where the narrow component is not related to the
NLR line profiles. This decomposition is inherently nonunique,
making it difficult to interpret their results.} (1996; shift
relative to \Hb). Richards et al. (2002) constructed composite
spectra for 794 SDSS quasars based on the \C\ versus \Mg\ velocity
shift, and also found a decreasing \C\ EW with increasing shift.
Here we confirm the \C\ EW versus shift correlation (\rS~=~0.376,
Table 2, note that the correlation sign is positive here since a
shift to the blue is defined as negative), but we note that the
\C\ EW shows significantly stronger correlations with other
parameters, in particular those related to the strength of
\Fe, \Ox\ and \aox. The \C\
shift shows comparable strength correlations with parameters
related to the strength of \Fe\ and \Ox.

Richards et al. (2002) found that the \C\ EW versus shift
correlation appears to be induced by absorption of the red wing of
\C, rather than a true shift of the line, which they suggested
(among other options) may imply that this correlation is induced
by inclination dependent
absorption effects, rather than physical changes in the BLR
properties. However, the strong correlations of the \C\ EW with some
of the BG92 EV1 parameters mentioned above, and the various
evidence that the EV1 correlations are driven by \Ledd\ (see also
BL04) argue that the correlation of the \C\ EW with the \C\ shift
is not just an inclination effect, but may rather be induced by
the dependence of both parameters on \Ledd.

\subsubsection{Correlations with \Fe~EW}

Marziani et al. (1996) find significant correlations of R~\Fe~EW
with R~\C~EW and \C~shift, which we verify here (Table 2),
although we find that R~\Fe~EW shows stronger correlations with
the \C~EW and with R~FWHM. As mentioned in BL04, the inverse
relation between the \Fe\ and \C\ strength may be an optical depth
effect, where large resonance line optical depth results in the
conversion of UV \Fe\ to optical \Fe\ (Netzer \& Wills 1983; Shang
et al. 2003), and the collisional destruction of the \C\ line
(Ferland et al. 1992). Alternatively, the \Fe\ and \C\ lines may
originate in different populations of clouds in the BLR, and the
inverse relation can be induced by the dominance of one population
at the expense of the other, e.g. due to obscuration (T. Boroson,
private communication).

In the `single population of clouds' interpretation the
differences between the \C\ and \Hb\ line profiles\footnote{The
\Hb\ line profiles are generally similar to \Fe\ (BG92).} may be
induced by a velocity dependent optical depth in the BLR. For
example, at higher velocities the line optical depth may drop,
causing enhancement of \C\ emission and suppression of the optical
\Fe. Alternatively, in the `two populations of clouds'
interpretation the difference in profiles may result from
different physical locations of the two populations. For example,
higher ionization lines may originate closer to the center, making
\C\ broader than \Hb. However, both interpretations do not explain
naturally why \C\ becomes {\em narrower} than \Hb\ when the
\Hb~FWHM~$>4000~$km~s$^{-1}$. A possible alternative
interpretation for the inverse trend in line widths is that \Hb\
and \C\ have orthogonal velocity fields. For example, \Hb\
originating in a high density low ionization Keplerian disk
component, and \C\ in a lower density higher ionization radial
outflow component (e.g. Wills et al. 1993a). The
\Hb~FWHM~$<4000$~km~s$^{-1}$ objects would then represent `face on
BLR disk' objects and \Hb~FWHM~$>4000$~km~s$^{-1}$ `edge on'
objects (e.g. Wills \& Browne 1986). If true, that would imply that
the \Hb\ based \Mbh\ estimates could be biased, but there is no
current supporting evidence for such a bias.
 Velocity resolved reverberation mapping will be
invaluable in disentangling the spatial distribution of the low
and high ionization line emission as a function of velocity in the
BLR (e.g. Horne et~al. 2004).

\subsection{Notable noncorrelations}

The correlation strength between the \C\ and \Hb\ shift, shape,
and asymmetry parameters are, respectively, 0.173, $-0.173$, and
0.136, which are all insignificant (see Table 3). This lack of
correlation is surprising. For example, if the \Hb\ and \C\ blue excess
asymmetry were produced by
radiation pressure effects, which may produce a large asymmetry in \C\
and a weaker one in \Hb, then one would expect the amount of line
asymmetry to be correlated as both would be driven by
\Ledd\footnote{Note that this mechanism may still be relevant for
the group of `I Zw 1-like' objects discussed above.}. The lack of
significant correlations between the \C\ and \Hb\ profile parameters
suggests that unrelated mechanisms influence the two line profiles.

The \C\ shift and shape parameters show at most marginally
significant correlations with the various optical and UV emission
parameters explored here. The \C\ asymmetry shows no significant
correlations at all (and was therefore not included in Table 2).
This, again, argues against the idea that the \C\ asymmetry is
generally induced by radiation pressure effects, which should have
produced a correlation with \Ledd\ and some of the EV1 parameters.
Corbin \& Boroson (1996) found a significant correlation
($-0.650$) between the \C\ asymmetry and the luminosity at
1549~\AA, however we find no significant correlation here
(\rS~$=-0.078$). Also, Sulentic et~al.
(2000) suggest there is a significant correlation between the \C\
shift/\C\ EW ratio and the \Hb\ FWHM, again this correlation
is not significant in our sample.

Leighly \& Moore (2004) find a significant correlation between the
\C\ asymmetry and \C\ EW  in a sample of NLS1s. Here we find in our
corresponding sub-sample of 17 narrow line PG AGN with
\Hb~FWHM~$<2000$~km~s$^{-1}$, a correlation of  0.314 (but only
0.007 for the complete sample) which is not significant. Leighly
(2004b) finds a significant correlation between the fraction of
the \C\ line flux at $v<$~0~km~s$^{-1}$ and \C\ EW and \aox\ in
their NLS1s sample. Here we find in our corresponding sample
correlations of $-$0.483 for the \C\ EW versus the \C\ line
flux at $v<$~0~km~s$^{-1}$, which is marginally significant, and
$-$0.053 for the correlation with \aox, which is clearly
insignificant.

\subsection{Indicators for dust absorption and scattering}

AGN display a range of optical-UV spectral slopes. This range
could be an intrinsic property of the continuum production
mechanism, or it may, at least in part, be induced by intrinsic dust
extinction. Below we present a set of correlations of various
parameters with \aouv, which suggest that dust has noticeable effects
on the continuum and line emission in our AGN sample, despite its
relatively blue color selection criterion.

Figure 9, upper panel, shows there is a significant correlation
between the \C/\Hb\ flux ratio and \aouv, such that AGN with a
redder continuum tend to have a relatively weaker \C\ emission. A
similar trend of weaker \C\ emission in redder quasars was noted
by Richards et al. (2003). The two diagonal lines in the upper
panel, which bound the distribution of most of the objects,
represent the effects of absorption by a foreground screen which
suppresses both the line and continuum emission by the same
factor\footnote{Note that \aouv\ is defined between 4861~\AA\ and
1549~\AA, the rest wavelengths of \Hb\ and \C.}. This correlation
appears to indicate that some of the range in the \C/\Hb\ flux
ratio is due to absorption by dust outside the BLR (as expected
due to sublimation, Laor \& Draine 1993), which reddens both the
lines and the continuum at a given wavelength by the same amount.
The intrinsic range in the \C/\Hb\ flux ratio may thus be closer
to a factor of $\sim 10$, rather than $\sim 30$ as Fig. 9
indicates. The \C/\Hb\ flux ratio also shows a strong correlation
with \aox\ (Table 2). Since the dust X-ray (2~keV) absorption is
negligible compared to the optical one (e.g. fig. 6 in Laor \&
Draine 1993), it cannot explain the correlation with \aox, which
may reflect the dependence of the intrinsic \C/\Hb\ line ratio on
the shape of the ionizing continuum (e.g. Korista et al. 1998).
Interestingly, the same trends were already noted by Netzer et al.
(1995) for the Ly$\alpha$/\Hb\ line ratio, which they find is also
strongly correlated with both \aouv\ and \aox, and which they
interpret as above.

The second panel shows there is a significant correlation of the
visible (white) light percentage polarization (Berriman et al. 1990) and
\aouv, such that the highest polarization increases as the objects
become redder. A plausible interpretation is that redder objects
tend to have a larger covering factor of dust, which will scatter
a larger fraction of the continuum light and produce a stronger
component of polarized scattered light. The increased extinction
of the transmitted, presumably unpolarized light, in steep
\aouv\ objects also
increases the relative contribution of the polarized light, and
thus the percentage polarization. The
presence of low polarization objects at the steepest \aouv\ may be
produced by geometrical dilution of the polarization amplitude, as
the polarization tends to zero towards a face-on view of an
azimuthally symmetric scattering source. Strongly reddened and
polarized continua are commonly observed in far IR selected AGN
(e.g. Wills 1999; Antonucci 2002; Hines et al. 2001), and the
above correlation suggests it extends with a smaller amplitude
to blue color selected AGN, such as the PG quasar sample.

The third panel shows there is a significant correlation between
\aouv\ and the \C\ absorption EW (from Laor \& Brandt 2002), such
that redder objects tend to have stronger \C\ absorption. This result
is consistent with the finding of
Constantin \& Shields (2003) that NLS1s with UV
absorption lines are on average redder than those without
absorption, and the finding of Hopkins et al. (2004) that intrinsic narrow
UV absorption lines are more common in redder quasars.
An association between UV absorption and reddening was
established by Sprayberry \& Foltz (1992) for low ionization broad
absorption line quasars (BALQs), and was later shown by Brotherton
et al. (2001) to extend with a smaller amplitude to all BALQs. Our
results suggest the UV absorption - reddening association is
present already at a UV absorption EW level of $\sim$~1-10~\AA.

A plausible interpretation for this correlation is that the \C\
absorption arises in gas associated with the dust which produces
the reddening (e.g. Crenshaw \& Kraemer 2001). Alternatively, the
dust and \C\ absorbers may be spatially distinct but coplanar. For
example, the dust absorption may originate in a torus like
structure, while the \C\ absorption may arise from an equatorial
wind launched from an accretion disk which is coplanar with a
larger scale torus. We note in passing that the \C\ absorption is
also correlated with \aox\ (Brandt et~al. 2000), which may
indicate the presence of an X-ray absorbing component coplanar
with the \C\ absorber. Since the \C\ emission strength is also
correlated with \aox, the X-ray absorber is likely to exist inside
the BLR, in order to affect the ionizing continuum incident on the
BLR.

Evidence for an association between reddening, UV absorption, and
a planar structure is seen in radio galaxies (Baker \& Hunstead
1995; Barthel, Tytler \& Vestergaard 1997; Baker et al. 2002;
Vestergaard 2003). A similar picture appears to hold for broad
absorption line quasars, which apart from the redder continuum
mentioned above, also show a higher polarization (e.g. Schmidt \&
Hines 1999; Ogle et al. 1999), and a weaker \C\ emission (Turnshek
1984; Hartig \& Baldwin 1986). The correlations shown in Fig.9
suggest that this picture may also extend to optically blue
selected quasars.

\begin{figure}
\includegraphics[width=120mm,angle=-90]{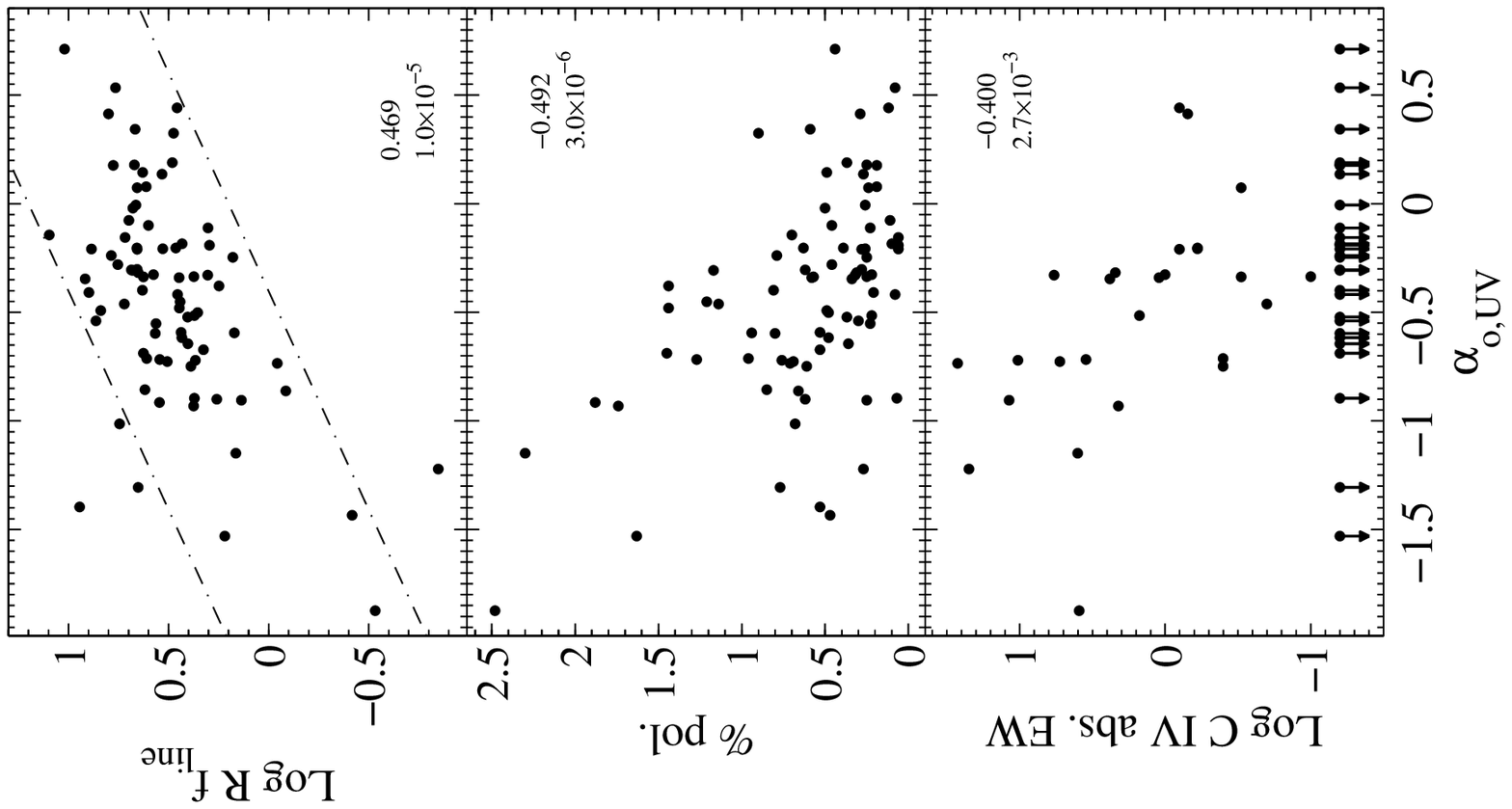}
\caption{Indications for dust absorption and scattering. Upper
panel: the relation of the \C/\Hb\ flux ratio and \aouv. The
dashed lines indicate the effects of absorption of the lines and
the continuum by the same amount at a given wavelength. The
reddening of the continuum and weakening of \C\ may be induced by
a foreground dust screen. Middle panel: the relation of the white
light polarization and \aouv. This may be induced by a
contribution of dust scattered and polarized light, which
increases in relative strength with increasing dust covering
factor and optical depth. Lower panel: the relation of the \C\
absorption EW (measured for 54 objects) and \aouv. The \C\
absorption and continuum reddening may arise in a foreground
dusty gas, or it could
arise in coplanar but spatially distinct components.}
\end{figure}

\section{Conclusions}
A study of the \C\ and \Hb\ emission line profiles in a nearly complete
sample of 81 low $z$ optically selected quasars reveals the following:

1. Narrow \C\ lines (FWHM~$< 2000$~km~s$^{-1}$) are rare ($\sim 2$
per cent occurrence rate) compared to narrow \Hb\ ($\sim 20$ per
cent).

2. When the \Hb\ FWHM~$<4000$~km~s$^{-1}$ the \C\ line is broader
than \Hb, but the reverse is true when the \Hb\
FWHM~$>4000$~km~s$^{-1}$. This argues against the view that
\C\ generally originates further inward in the BLR
compared with \Hb.

3. \C\ appears to provide a significantly less accurate, and
possibly biased estimate
of the black hole mass in AGN, compared with \Hb.

4. The line profile differences are correlated with some of the
BG92 eigenvector 1 parameters, which may be related to the
relative accretion rate \Ledd. A high \Ledd\ appears to be a
necessary, but somehow not sufficient condition for having
strongly blueshifted and asymmetric \C\ emission.

 Thus, a simple understanding of what controls the \C\ profile and the large range
in \C\ to \Hb\ FWHM ratio remains a challenge.

5. There are indications for dust reddening and scattering in
`normal' optically selected AGN. In particular, PG quasars with a
redder optical-UV continuum slope show weaker \C\ emission,
stronger \C\ absorption, and a higher continuum polarization.

Reverberation mappings of AGN with highly discrepant \Hb\ and
\C\ profiles can provide important constraints on the spatial
distribution of the line emissivity as a function of velocity,
and thus test the various scenarios for the origin of the
large profile differences.

\section*{acknowledgments}
We thank T. Boroson for providing the optical spectra and accurate
redshifts for all objects, and for helpful discussions. We also
thank Marianne Vestergaard, Gordon Richards, and the referee
Craig Warner for careful reading of the manuscript and many helpful
comments. This
research has made use of the NASA/IPAC Extragalactic Database
(NED), which is operated by the Jet Propulsion Laboratory,
California Institute of Technology, under contract with the
National Aeronautics and Space Administration.

\end{document}